\documentclass[preprint,preprintnumbers,amsmath,amssymb]{revtex4}

\usepackage{graphicx}% Include figure files
\usepackage{dcolumn}% Align table columns on decimal point
\usepackage{bm}% bold math

\def\fun#1#2{\lower3.6pt\vbox{\baselineskip0pt\lineskip.9pt
\ialign{$\mathsurround=0pt#1\hfil##\hfil$\crcr#2\crcr\sim\crcr}}}

\newcommand\lsim{\mathrel{\rlap{\lower4pt\hbox{\hskip1pt$\sim$}}
    \raise1pt\hbox{$<$}}}
\newcommand\gsim{\mathrel{\rlap{\lower4pt\hbox{\hskip1pt$\sim$}}
    \raise1pt\hbox{$>$}}}
\def\lsim{\mathrel{\raise.3ex\hbox{$<$\kern-.75em\lower1ex\hbox{$\sim$}}}} 
\def\gsim{\mathrel{\raise.3ex\hbox{$>$\kern-.75em\lower1ex\hbox{$\sim$}}}}

\newcommand{\mchi}{m_{\chi}}

\newcommand{\svave}{\langle\sigma v\rangle}

\newcommand{\tev}{\,\mbox{TeV}}
\newcommand{\gev}{\,\mbox{GeV}}

\begin{document}

\title{Dark Matter at the Centers of Galaxies \\ David Merritt \\
Chapter 5 from {\it Particle Dark Matter: Observations, Models and Searches}\\ 
Ed. G. Bertone (Cambridge University Press)}
%\author{David Merritt}
\bigskip
\maketitle

Dark matter halos formed in $\Lambda$CDM cosmologies exhibit a
characteristic dependence of density on distance from the center.
%\cite{Chap:Moore}.
Early studies \cite{DC-91,NFW-96} established 
$\rho_{\rm DM}\sim r^{-3}$ or $r^{-4}$ at large radii and
$\rho_{\rm DM}\sim r^{-1}$ inside the virial radius.
On still smaller scales, the form of $\rho_{\rm DM}(r)$
was little more than an ansatz since the relevant scales were 
barely resolved in the $N$-body simulations.
A debate ensued as to whether the profiles were indeed universal, 
and if so, what power of the radius described the dark matter density 
in the limit $r\rightarrow 0$.
Subsequent studies found both steeper 
\cite{ghigna-98,moore-98,klypin-01,diemand-04} 
and shallower \cite{navarro-04,merritt-05,navarro-08}
central profiles.

The focus of this chapter is the dark matter distribution on
 sub-parsec scales.
At these radii, the gravitational force in many galaxies 
is known to be dominated by the observed
baryonic components (stellar bulge, nuclear star cluster)
and by the supermassive black hole.
Dark matter densities at these radii are barely constrained
observationally; however they could plausibly be orders of magnitude
higher than the local value at the Solar circle
($\sim 10^{-2} M_\odot$ pc$^{-3}$), due both to the special location
at the center of the halo, and also to interactions between dark matter 
and baryons during and after formation of the galaxy.
High dark matter densities make the centers of galaxies
preferred targets for {\it indirect detection} studies, in which
secondary particles and photons from the annihilation or decay of
supersymmetric dark matter particles are detected on the Earth.
%\cite{Chap:Bergstrom}.

\section{Phenomenology of Galactic Nuclei}
\label{sec:phenom}
The distribution of {\it baryonic} matter at the centers of galaxies
is relevant to the dark matter problem for several reasons.
\begin{itemize}
\item Many dynamical processes affect the dark and luminous
components in similar ways. The distribution of stars at the center
of a galaxy can tell us something about the distribution of 
dark matter.
\item If the nuclear relaxation time (eq.~\ref{eq:tr})
is shorter than  the age of the universe,
stars will exchange kinetic energy with dark matter particles causing 
the dark matter distribution to evolve in predictable ways.
\item Supermassive black holes appear to be ubiquitous components
of galactic nuclei. Depending on its mode of growth,
a SMBH can greatly increase, or decrease, the density of dark
matter in its vicinity.
\end{itemize}

Galactic nuclei are the innermost regions of stellar {\it spheroids}:
either elliptical galaxies, or the bulges of spiral galaxies.
Most galaxies are too distant for individual stars to be resolved,
and descriptions of their structure are generally based on their
{\it luminosity profiles}, the surface brightness of starlight 
as a function of distance from the center.
Luminosity profiles of galactic spheroids are well fit at most radii 
by S\'ersic's \cite{sersic-68} law,
\begin{equation}
\ln I = \ln I_0 - bR^{1/n}
\label{eq:sersic}
\end{equation}
where $I$ is the surface brightness at projected radius $R$ and
$n$ is the S\'ersic index; $n=4$ is the de Vaucouleurs law \cite{devauc-48}.
S\'ersic's law predicts a space density that increases as
$\rho_\star\sim r^{-(n-1)/n}$ toward the center, 
or $\rho_\star\sim r^{-3/4}$ in
the case of de Vaucouleurs' law.
However in the best-resolved galaxies -- which include both the nearest, 
and the intrinsically largest, galaxies -- 
deviations from S\'ersic's law  often appear near the center.
Bright spheroids exhibit deficits with respect
to S\'ersic's law, or {\it cores}.
Faint spheroids exhibit excesses, or {\it nuclear star clusters}
(NSCs), with sizes in the range $1-100$ pc and luminosities
in the range $10^6-10^8 L_\odot$ \cite{boeker-02,ACS8}.
The transition from deficit to excess occurs at spheroid luminosities
of roughly $10^{10}L_\odot$ \cite{ACSFornax2}.
While NSCs are generally unresolved (a notable exception  \cite{schoedel-08} 
being the NSC at the center of the Milky Way),
cores in luminous elliptical galaxies can extend hundreds of parsecs.

Galactic nuclei also contain supermassive black holes (SMBHs).
In a handful of galaxies, the presence of the SMBH is indicated
by a clear Keplerian rise of stellar or gas velocities
inside a radius $\sim r_h$, 
the {\it gravitational influence radius}:
\begin{equation}
r_h = GM_\bullet/\sigma^2.
\label{eq:rh}
\end{equation}
Here $M_\bullet$ is the black hole mass and $\sigma$ is the
one-dimensional, rms velocity of stars in the spheroid.
In other galaxies, indications are seen of a central
rise in velocity but the implied SMBH mass is very uncertain
\cite{MF-01c}.
Among the $\sim$ dozen galaxies with well-determined SMBH masses,
there is a remarkably tight correlation between $M_\bullet$ and
$\sigma$, the $M_\bullet-\sigma$ {\it relation} \cite{FF-05}:
\begin{equation}
M_\bullet/10^8M_\odot \approx 1.66 (\sigma/200\ {\rm km\ s}^{-1})^\alpha, \ \ \ \ \alpha\approx 4.86 .
\label{eq:ms}
\end{equation}
Combining equations (\ref{eq:rh}) and (\ref{eq:ms}),
\begin{equation}
r_h \approx 18\ {\rm pc}\ (\sigma/200\ {\rm km\ s}^{-1})^{2.86}
\approx 13\ {\rm pc}\ (M_\bullet/10^8 M_\odot)^{0.59}.
\end{equation}
The $M_\bullet-\sigma$ relation extends at least down to 
$M_\bullet\approx 10^{6.6} M_\odot$,
the mass of the Milky Way SMBH \cite{gillessen-08,ghez-08}.
Indirect, but contested, evidence exists for lower mass,
{\it intermediate-mass black holes}  
in some low-luminosity spheroids, active galaxies, 
and star clusters \cite{MC-04}.

The connection between SMBHs and nuclear structure 
is circumstantial but reasonably compelling.
Observed core radii are $\sim$ a few $r_h$ in the brightest elliptical
galaxies, consistent with a model in which
the cores were created when stars were displaced by a pre-existing 
binary SMBH \cite{merritt-06}.
At the other extreme in spheroid luminosity, NSCs appear to sometimes co-exist
with SMBHs, but there are only a handful of galaxies in which the
presence of both components can unambiguously be established \cite{seth-08};
thus there is no clear evidence that SMBHs are associated with
an excess of (luminous) mass at the centers of galaxies.

The nuclear {\it relaxation time}
\begin{equation}
t_r = {0.34\sigma^3\over G^2 \rho_\star m_\star \ln\Lambda}
\label{eq:tr}
\end{equation}
measures the time scale over which gravitational encounters 
redistribute energy between stars; $\ln\Lambda\approx 12$ is
the Coulomb logarithm \cite{spitzer-87}.
Relaxation times greatly exceed 10 Gyr at all radii in spheroids
more massive than $\sim 10^{10}M_\odot$ \cite{MMS-07}; 
in these ``collisionless'' systems, star-star and star-dark matter
interactions occur too rarely to significantly alter the distribution
of either component 
over the lifetime of the galaxy.
In fainter spheroids, and particularly those containing dense NSCs, 
central relaxation times can be shorter 
\cite{merritt-09};
for instance, at the Galactic center, $t_r $ falls below 10 Gyr
inside $r_h$ \cite{schoedel-07}.
In these ``collisional'' nuclei, the distribution of stars 
around a SMBH is expected to evolve, in a time $\sim t_r$,
to the quasi-steady-state form
\begin{equation}
\rho_\star(r) \propto r^{-7/4}
\label{eq:bw}
\end{equation}
at $r\lesssim r_h$: a {\it Bahcall-Wolf cusp} \cite{BW-76}.
If multiple mass groups are present, equation~(\ref{eq:bw})
describes the central behavior of the most massive component,
while the lowest-mass component (e.g. dark matter particles)
obeys
\begin{equation}
\rho(r)\propto r^{-3/2}
\label{eq:mmbw}
\end{equation}
 \cite{BW-77}.
Equation~(\ref{eq:bw}) approximately describes the
distribution of luminous stars at the Galactic center \cite{schoedel-07}, but
no other galaxy containing a SMBH is near enough that a
Bahcall-Wolf cusp could be resolved even if present.

\begin{figure}
\centering\leavevmode
\mbox{
\includegraphics[width=1.9in,angle=-90.]{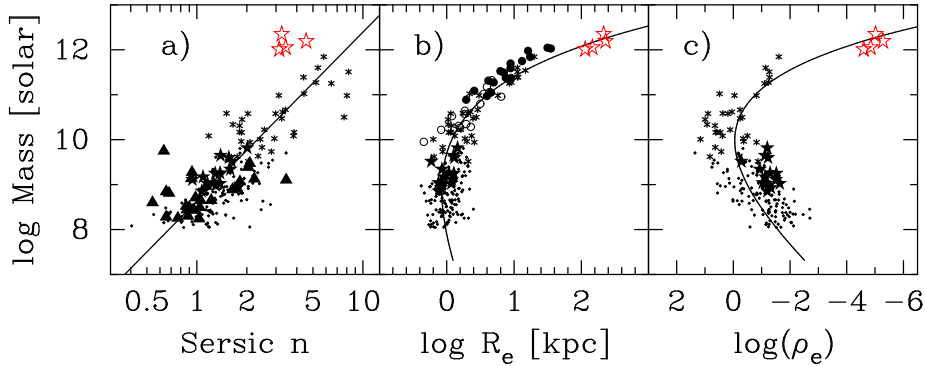}
}
\caption{Relations between the parameters that define the
mass, size and density of galaxy spheroids and $N$-body
dark matter halos.
Because both types of system are well described by the same
empirical density law, they can be plotted together on the same
axes.
$n = $ S\'ersic index, $R_e = $ effective (projected half-mass or
half-light) radius, $\rho_e = $ mass density at $r=R_e$.
The filled symbols are dwarf and giant elliptical galaxies;
structural parameters for these objects were derived by fitting
S\'ersic laws, eq.~(\ref{eq:sersic}),  to their observed luminosity profiles.
Luminosity densities were converted to mass densities  using an assumed 
value for the stellar mass-to-light ratio, neglecting dark matter.
The open (red) stars are a set of simulated, galaxy-sized dark matter halos
\cite{diemand-04}.
Mass density profiles for these objects were fit with de-projected S\'ersic
laws and total mass was defined as mass within the virial radius.
The solid lines are semi-empirical fitting relations.
(Adapted from \cite{empirical3}.)
}
\label{fig:merritt_1}
\end{figure}

Scaling relations between spheroid luminosities and masses or
velocity dispersions are continuous over many decades
in mass, from giant elliptical galaxies down to globular 
clusters (Fig.~\ref{fig:merritt_1}).
Only the class of dwarf spheroidal galaxies (dSphs) depart
systematically from these relations, in the sense of having too large an
inferred (dynamical) mass compared with their luminosities:
these systems appear to be dark-matter dominated even at their centers
(\S~\ref{ssec:dsph}).

\section{Dark Matter Models}
\subsection{Cusps vs. cores}

Traditionally there have been two approaches
to estimating the density of dark matter at 
the centers of galaxies.
Unfortunately, they often lead to different conclusions.

{\it $N$-body simulations of gravitational clustering}
follow the growth of dark matter halos as they evolve 
via mergers in an expanding, cold-dark-matter universe.
Halo density profiles in these simulations are well determined 
on scales $10^{-2}\lesssim r/r_{vir} \lesssim 10^0$,
where the virial radius $r_{vir}$ is of order
$10^2$ kpc for a galaxy like the Milky Way; 
hence inferences about the dark matter density on parsec or sub-parsec scales
require an extrapolation from the $N$-body results.
A standard parametrization of $\rho_{\rm DM}$ in these
simulated halos is
\begin{equation}
\rho_{\rm DM}(r) = \rho_0\xi^{-1}\left(1+\xi\right)^{-2}
\label{eq:NFW}
\end{equation}
\cite{NFW-96}, the {\it NFW profile},
where $\xi=r/r_s$ and $r_s$ is a scale length
of order $r_{vir}$.
In the Milky Way, $r_{vir}\gg R_\odot$ (the radius of the
Solar circle) hence eq.~(\ref{eq:NFW}) is essentially a power law 
at $r<R_\odot$ and the implied dark matter density is
\begin{equation}
\rho_{\rm DM}(r)\approx 10^2 M_\odot {\rm pc}^{-3} 
\left({\rho_\odot\over 10^{-2}M_\odot {\rm pc}^{-3}}\right)
\left({R_\odot\over 8\ {\rm kpc}}\right)
\left({r\over 1\ {\rm pc}}\right)^{-1}
\label{eq:NFW2}
\end{equation}
where $\rho_\odot\equiv\rho_{\rm DM}(R_\odot)$ and
$\rho_\odot\approx 8\times 10^{-3}M_\odot {\rm pc}^{-3}$
(from the Galactic rotation curve).
%Based on halo scaling relations 
%\cite{Empirical3},
%DM densities at $r_h$ would be higher in spheroids fainter
%than that of the Milky Way.

{\it Rotation-curve studies of low-surface-brightness spiral
galaxies}
are generally interpreted as implying much lower, central
dark matter densities
\cite{Burkert-95,SB-00,Blok-02,Gentile-05,Blok-05}.
While there are caveats to this interpretation --
systematic biases in long-slit observations \cite{SGH-05},
non-circular motions \cite{Simon-05},
gas pressure \cite{valenzuela-07}, etc. --
these effects do not seem capable of fully explaining  the
discrepancies between rotation curve data and 
expressions like (\ref{eq:NFW}) \cite{Blok-04,Gentile-05}.
A model for $\rho_{\rm DM}(r)$ that is often fit to rotation
curve data is
\begin{equation}
\rho_{\rm DM}(r) = \rho_c\left(1+\xi\right)^{-1}\left(1+\xi^2\right)^{-1},
\label{eq:Burkert}
\end{equation}
the {\it Burkert profile} \cite{Burkert-95}, where $\xi\equiv r/r_c$ and
$r_c$ is the core radius.
Inferred core radii are $\sim 10^2-10^3$ pc and inferred
central densities are
$10^{-2}\lesssim \rho_c \lesssim 10^{0} M_\odot {\rm pc}^{-3}$.

Since the $N$-body halos are not resolved on the scales
($\sim 10^2$ pc) where rotation curves are measured,
the mismatch between theory and observation may be due in part 
to a poor choice of empirical function used to 
describe the $N$-body models.
An alternative parametrization
\begin{equation}
\rho_{\rm DM}(r) = \rho_0 \exp\left[-\left(r/r_0\right)^{1/n}\right],
\label{eq:einasto}
\end{equation}
the {\it Einasto profile} \cite{einasto-65},
has recently been shown to describe $N$-body haloes
even better than eq.~(\ref{eq:NFW})
\cite{navarro-04,merritt-06,prada-06}.
The low central density of the Einasto model alleviates some, but not all,
of the disagreement with rotation curve studies \cite{empirical3}.

Remarkably, eq.~(\ref{eq:einasto}) has the same functional
form as S\'ersic's law (\ref{eq:sersic}) that describes
 the {\it projected} density profiles of
galactic spheroids.
In fact, the two descriptions are roughly equivalent
if $n_{\rm Einasto} \approx n_{\rm Sersic}+1$
\cite{merritt-05}, showing that luminous spheroids and
simulated dark-matter halos are essentially rescaled
versions of each other (Fig.~\ref{fig:merritt_1}),
at least over the range in radii that is resolvable by the 
$N$-body simulations.

Several resolutions have been suggested for the persistent conflict
between predicted and measured, central dark matter densities
\cite{silk-02,tasi-03,SM-02},
but none is universally agreed upon. 

\subsection{Effects of baryonic dissipation}
\label{ssec:adiabat}

$N$-body simulations of dark matter clustering typically ignore the
influence of the baryons (stars, gas) even though these
components may dominate the gravitational force in the 
inner kiloparsec or so
One simple, though idealized, 
way to account for the effect of the baryons on the dark matter
is via {\it adiabatic contraction}  models,
which posit that the baryons contracted quasi-statically and
symmetrically within the pre-existing dark matter halo, 
pulling in the dark matter and increasing its density in the process
\cite{blumenthal-86}.
When applied to a dark matter halo with the density law
(\ref{eq:NFW}), i.e. $\rho_{\rm DM}\sim r^{-1}$,
the result is the more steeply rising $\rho\sim r^{-\gamma_c}$,
$\gamma_c\approx 1.5$ \cite{prada-04,gnedin-04,mambrini-06}.

\begin{figure}
\centering
\includegraphics[width=3.5in,angle=0]{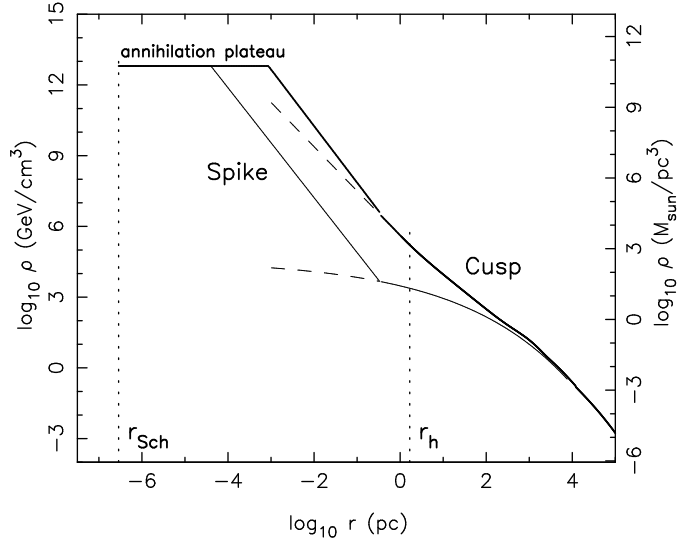}
\vspace{0cm}
\caption{Possible models for the dark matter distribution at
the center of a galaxy like the Milky Way.
The thin curve shows an Einasto density profile,
eq.~(\ref{eq:einasto}),
and the thick curve labelled ``cusp'' is the same model after 
``adiabatic compression'' by the baryons (stars and gas).
Lines labelled ``spike'' show the additional increase in density that
would  result from spherically-symmetric growth of the SMBH.
The ``annihilation plateau'' is the density that satisfies
$\rho=\mchi/\svave t$; this density was computed 
assuming $\mchi=200$ \gev,
$\svave=10^{-28}$ cm$^3$ s$^{-1}$ and $t=10^{10}$ yr.
Dashed vertical lines indicate the SMBH's Schwarzschild
radius (left) and gravitational influence radius (right).
Effects of the dynamical processes discussed in \S\ref{sec:dynam}
(scattering of dark matter off stars, loss of dark matter
into  the SMBH) are excluded from this plot;
in nuclei (like that of the Milky Way) with relaxation times less 
than $\sim 10$ Gyr,
these processes would generally act to decrease the dark matter
density below what is shown here, particularly in the models
with a spike (cf. Fig.~\ref{fig:merritt_4},\ref{fig:merritt_5}).
(Adapted from \cite{BM-05}.)}
\label{fig:merritt_2}
\end{figure}

\section{Dark Matter in Collisionless Nuclei}

The dark matter annihilation signal from a region of volume
$V$ is proportional to $\langle\rho_{\rm DM}^2\rangle V$.
If the dark matter density rises steeply toward the center
of a galaxy, the annihilation flux can be dominated by
dark matter within the central parsec or so.
Neither $N$-body simulations, nor rotation curve studies,
are a reliable guide to $\rho_{\rm DM}$ on these small
scales.
In addition, in many galaxies, the total gravitational force 
in the inner parsecs is dominated by the SMBH.

We consider first ``collisionless'' nuclei, in which central
relaxation times exceed $\sim 10$ Gyr; this is the case in
spheroids more massive than $\sim 10^{10}M_\odot$ \cite{MMS-07}.
In these systems, the distribution of stars and dark matter 
near the galaxy center has probably remained essentially unchanged
since the era at which the nucleus and the SMBH were created.

\subsection{Black hole adiabatic growth models}

If the SMBH grew to its final size in the simplest possible way --
via spherically-symmetric infall of gas -- the density
of matter around it would increase \cite{peebles-72,young-80}, 
in the same way that contracting baryons steepen the dark matter 
density profile on somewhat larger scales (\S~\ref{ssec:adiabat}). 
In the limit that the growth timescale of the SMBH is long 
compared with orbital periods, this scenario predicts a final 
density (of stars or dark matter) near the SMBH of
\begin{equation}
\rho_f(r) \approx \rho_i(r_f)(r/ r_f)^{-\gamma},
\ \ \ \ \gamma_f=2 + 1/(4-\gamma_i)
\label{eq:adiabat}
\end{equation}
where $\rho_i\propto r^{-\gamma_i}$ is the pre-existing density profile,
and $r_f\approx 0.2 r_h$.
Even for  $\gamma_i\approx 0$, eq.~(\ref{eq:adiabat}) predicts
$\gamma_f> 2$ -- a density 
{\it spike} (Fig.~\ref{fig:merritt_2}).
Such a steep dark matter density profile near the SMBH would imply 
very high rates of dark matter annihilation \cite{GS-99}.

Stars would respond in the same way as dark matter particles to the growth
of a SMBH.
A $\rho\sim r^{-2}$ density cusp in the stars is not observed at $r<r_h$
in any galaxy however, even those close enough that a spike could be
resolved if present.
In the case of the most luminous galaxies, this is an expected consequence
of core formation by binary SMBHs, as discussed in the next section.
In low-luminosity spheroids like the bulge of the Milky Way,
relaxation times are short enough to convert a stellar spike into a shallower,
Bahcall-Wolf cusp in $\sim$ one relaxation time.

It is also possible that spikes never form.
Even small (compared with $r_h$) and temporary displacements of the SMBH 
from its central location are sufficient to inhibit the growth of a spike 
or to destroy it after it has formed \cite{ullio-01}.
Most models for the growth of SMBHs 
invoke {\it strong} departures
from spherical symmetry during galaxy mergers in order 
to remove excess angular momentum
from the infalling gas \cite{shlosman-89}.

\begin{figure}[!b]
\label{fig:2}
\begin{minipage}{0.59\textwidth}
\includegraphics[width=0.9\textwidth]{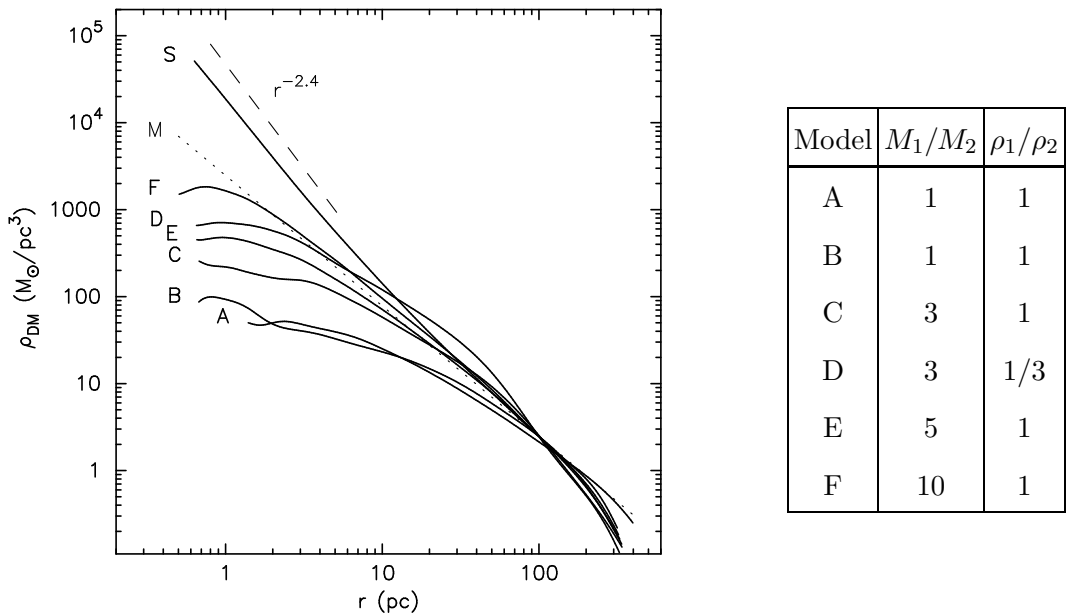}
\end{minipage}
\begin{minipage}{0.35\textwidth}
\begin{tabular}{|c|c|c|}
\hline
Model & $M_1/M_2$ & $\rho_1/\rho_2$
\\ 
\hline
A & 1  & 1    \\
B & 1  & 1    \\
C & 3  & 1    \\
D & 3  & 1/3  \\
E & 5  & 1    \\
F & 10 & 1    \\
\hline
\end{tabular}
\end{minipage}
\caption{The  effect of mergers, including SMBHs, on the central
densities of dark matter halos \cite{MMVJ-02}.
The curves labelled $M$ and $S$ are the density profiles of the
larger of the two haloes before and after a SMBH was grown
adiabatically at its center.
The other curves show the final density profile of the merged
halo, for various choices of the initial halo parameters, as given
in the table.
$M_1$ ($M_2$) is the mass of the large (small) dark matter halo and
$\rho$ is the central halo density before growth of the SMBH.
Mass and length scalings were based on the Milky Way.}
\label{fig:merritt_3}
\end{figure}

\subsection{Binary black holes and core creation}

Most spheroids are believed to have experienced at least one
major merger (defined as a merger with mass ratio $\sim 3:1$ or less)
since the epoch at which the SMBHs formed \cite{HK-00}.
If two merging galaxies each contain a SMBH, a massive binary
forms \cite{BBR-80}, displacing stars and dark matter as the two
holes spiral in to the center.
The process can be understood as a kind of dynamical friction, with
the ``heavy particles'' (the SMBHs) transferring their kinetic energy
to the ``light particles'' (stars, dark matter). 
However, most of the energy transfer takes place after the two SMBHs
have come within each other's spheres of influence, and in this regime
the interaction with  the background is dominated by another mechanism,
the {\it gravitational slingshot} \cite{SVA-74}.
The massive binary ejects passing stars or dark matter particles
at high velocity, removing them from the nucleus and simultaneously
increasing its binding energy \cite{quinlan-96}.

This process stops, or at least slows, when the two SMBHs reach
a separation $r_{\rm stall}\approx q/(1+q)^2 r_h$, 
the {\it stalling radius}; here $q\equiv m_2/m_1 \le 1$ is the binary
mass ratio.
At this separation, the binary has already removed essentially all 
material on intersecting orbits and the inspiral stops; or, it continues
at a much lower rate that is limited by how fast the depleted orbits
can be repopulated \cite{MMS-07}.
The size of the low-density core that is produced by inspiral
from $r\approx r_h$ to $r\approx r_{\rm stall}$ is 
$\sim$ a few times $r_h$, quite consistent with the 
sizes of the stellar cores observed in many galaxies 
\cite{graham-04,merritt-06}. 
Dark matter cores would presumably be of similar size 
(Fig.~\ref{fig:merritt_3}), 
or even larger if multiple mergers occurred \cite{volonteri-03,merritt-06}
or if the pre-binary dark matter distribution was 
characterized by a core as in the Burkert model described above. 
This mechanism can probably {\it not} explain the kpc-scale dark matter
cores inferred in many spiral galaxies, however:
the mergers that formed the bulges of these systems would
have resulted in much smaller, parsec-scale cores.

\subsection{Gravitational-wave recoil}
If the two SMBHs at the center of a merged galaxy manage to overcome 
the ``final-parsec problem'' and coalesce, another mechanism comes into 
play that can affect the central density of stars and dark matter.
Emission of gravitational waves during the final plunge is
generically anisotropic, resulting in a transfer of linear
momentum to the coalesced SMBH \cite{RR-89}.
The resultant ``kick'' can be as large as $\sim 4000$ km s$^{-1}$ if
the two holes have equal mass and optimal spins
(i.e. maximal amplitude, oppositely aligned, 
and parallel to the binary orbital plane)
\cite{lousto-07}.
While such extreme kicks are probably rare, even a mass ratio of
$0.1$ can result in kicks of $\sim 1000$ km s$^{-1}$ if spins 
are optimal; while if the spins are maximal but oriented parallel 
to the orbital angular momentum, 
the kick velocity peaks at $\sim 600$ km s$^{-1}$ for $m_1=m_2$.
By comparison, kicks large enough to remove SMBHs from galaxy cores range from
$\sim 90$ km s$^{-1}$ for spheroid masses of $3\times 10^9M_\odot$ to 
$\sim 750$ km s$^{-1}$ for $M_{\rm sph}=3\times 10^{11}M_\odot$ to 
$\sim 1000$ km s$^{-1}$ for $M_{\rm sph}=3\times 10^{12}M_\odot$
\cite{mmfhh-04}.

Sudden removal of the SMBH from the galaxy center impulsively
reduces the force that binds stars and dark matter to the center
\cite{mmfhh-04,copycats-04}.
If $V_{\rm kick}$ is less than the galaxy central escape velocity,
still more energy is injected into the core by the kicked SMBH as
it passes repeatedly through the center before finally coming to rest.
Cores enlarged in this way can be several times larger than $r_h$,
and indeed a few of the brightest elliptical galaxies have such
over-sized cores \cite{lauer-07}; 
dark matter cores are presumably of comparable size in these galaxies.

\section{Dark Matter in Collisional Nuclei}
\label{sec:dynam}

Nuclear relaxation times fall below $10$ Gyr in spheroids fainter
than $\sim 10^{10}L_\odot$, roughly the luminosity where NSCs first appear
\cite{MMS-07}.
As discussed above, at least some of these galaxies (including
the Milky Way) also contain SMBHs.
In these collisional nuclei, a Bahcall-Wolf cusp in the stars can re-form
even if it had been previously destroyed by a binary SMBH \cite{MS-06}.
Both the Milky Way and the nearby dwarf elliptical galaxy M32
exhibit steeply-rising stellar density profiles 
within the influence radii ($r\lesssim 1$ pc) of their SMBHs
\cite{schoedel-08,lauer-98}.

Dark matter particles in these galaxies 
are still collisionless: their individual masses are so
small that gravitational encounters between them are negligible.
But even massless particles can scatter off of stars, and the
associated time scale is equal to within a factor of order unity
to the star-star relaxation time, eq.~(\ref{eq:tr})
\cite{ilyin-04,merritt-04,gnedin-04}.

Naively, one would expect the stars to act like a heat source,
transferring kinetic energy to the dark matter particles and
lowering their density.
This does occur; but in addition, the phase-space density of
dark matter particles is driven toward a constant value as a function
of orbital energy, $f(E)\approx f_0$.
A constant phase-space density with respect to $E$ 
implies a configuration-space density that rises as $\rho\sim r^{-3/2}$
in the $1/r$ potential of a SMBH.
The term {\it crest}, for ``collisionally-regenerated structure''
has been coined to describe the result of this process \cite{MHB-07}.

\begin{figure}[t]
\centering\leavevmode
\mbox{
\includegraphics[width=2.4in,angle=-90.]{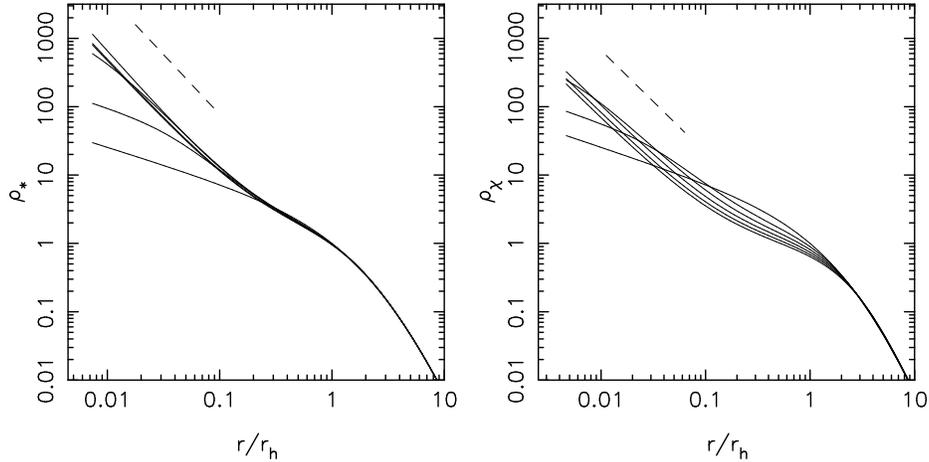}
}
\caption{Joint evolution of the stellar (left) and dark-matter 
(right) densities around a SMBH due to star-star and
star-dark matter gravitational encounters \cite{MHB-07}.
Length unit $r_h$ is the SMBH influence radius
(roughly 1 pc at the Galactic center).
Density is in units of its initial value at $r_h$.
Curves show density profiles
at times $(0,0.2,0.4,0.6,0.8,1.0)$ in units of the initial
relaxation time (Eq.~\ref{eq:tr}) at $r=r_h$.
Dashed lines are the ``steady-state'' solutions,
eqs.~(\ref{eq:bw}) and (\ref{eq:mmbw}).}
\label{fig:merritt_4}
\end{figure}

Fig.~\ref{fig:merritt_4} illustrates the joint evolution of the stellar
and dark-matter
densities near a SMBH at the center of a galaxy in which the
density of both components had previously been lowered by a
binary SMBH.
The stars are seen to attain the $\rho\sim r^{-7/4}$, Bahcall-Wolf
form in approximately one relaxation time.
Simultaneously, the dark matter particles evolve
to the shallower $\rho\sim r^{-3/2}$ profile, increasing
their density near the SMBH.
The normalization of the dark matter 
density continuously drops
as the stars transfer kinetic energy to the dark matter.
Simulations like these suggest that the presence of a
Bahcall-Wolf (collisional) cusp in the stars at the
center of a galaxy should always be associated with a 
shallower, $\sim r^{-3/2}$ ``crest'' in the dark matter,
regardless of how the nucleus and the SMBH formed
\cite{MHB-07}.
Note, however, that this argument can not be used to infer the 
{\it normalization} of the dark matter density.

Conditions for the formation of crests are relaxed somewhat
if there is a top-heavy spectrum of stellar masses since
the dark matter scattering time scales as $\tilde m_\star^{-1}$ where
$\tilde m_\star = \langle m_\star^2\rangle/\langle m_\star\rangle$
is the second moment of the stellar mass function
\cite{merritt-04}.
The stellar cusp can also evolve more quickly in this case
\cite{baumgardt-04b}.
However, due to their gradual dissolution,
dark matter crests might only be present, with significant amplitudes, 
in galactic nuclei having a fairly narrow range of properties:
older than $\sim$one relaxation time but younger than many relaxation times.
The corresponding range in spheroid
luminosities is approximately
$3\times 10^8 L_\odot\lesssim L\lesssim 3\times 10^9 L_\odot$
\cite{MHB-07}.
In addition, low-luminosity spheroids may not all contain massive 
black  holes.
In the absence of a SMBH, the stars would undergo {\it core collapse},
in a time that shorter than $\lesssim 10^{10}$ yr in the densest nuclei
\cite{merritt-09},
producing a $r^{-2.25}$ stellar density profile;
the dark matter density would be expected to evolve only 
slightly in this case \cite{Kim-04}.

\section{The Galactic Center}

The proximity of the Galactic center makes it a promising target
for indirect detection studies: predicted fluxes can be more than
an order of magnitude higher than for any other potential
galactic source \cite{Stecker-88,BUB-98,BSS-01}.
In addition, observations of stellar velocities in the inner parsec 
of the Milky Way yield a highly precise value for the mass in the SMBH
\cite{gillessen-08,ghez-08}, as well as 
(somewhat less precise) estimates of the distributed mass
\cite{trippe-08,pmpaper-08}.
In principle, dark matter might be detected by observing its
effects on the stellar orbits \cite{hall-06,zakharov-08}, 
but uncertainties about the masses
associated with other ``dark'' components -- neutron stars,
stellar mass black holes, etc. -- probably render this approach
unfeasible for the forseeable future.

Given a detector with angular acceptance $\Delta\Omega$ sr,
the observed flux of photons produced by
annihilation of dark matter particles is \cite{BUB-98}
\begin{equation}
\Phi(\Delta\Omega,E)\approx 1.9\times 10^{-12}{dN\over dE} 
{\svave\over 10^{-26}{\rm cm}^{-3} {\rm s}^{-1}} 
\left({1{\rm TeV}\over \mchi}\right)^2
\overline{J}_{\Delta\Omega}\Delta\Omega\ {\rm cm}^{-2} {\rm s}^{-1}
\end{equation}
where $dN/dE$ is the spectrum of secondary photons per annihilation,
$\mchi$ is the particle mass,
$\svave$ is the velocity-averaged self-annihilation 
cross section, 
and $\overline{J}_{\Delta\Omega}$ contains the information about the
dark matter density:
\begin{equation}
\overline{J}_{\Delta\Omega} = K\Delta\Omega^{-1}\int_{\Delta\Omega} d\psi\int_\psi \rho_{\rm DM}^2 dl.
\label{eq:jdo}
\end{equation}
Here, $dl$ is an element of length along the line of sight
and $\psi$ is the angle with respect to the Galactic center.
The normalizing factor $K$ is typically set to
$K^{-1}=(8.5{\rm kpc})(0.3\gev/{\rm cm}^3)^2$:
the product of the distance to the Galactic center,
and the squared, local value of the dark matter density,
the latter derived from the measured rotation curve 
assuming an NFW halo.
Henceforth we write 
$\overline{J}_{\Delta\Omega = 10^{-5}}\equiv\overline{J}_5$:
$\Delta\Omega=10^{-5}$ sr ($\sim 10$ arc minutes)
is the approximate angular resolution of
atmospheric Cerenkov telescopes
like H.E.S.S. \citep{HESS-00} and of FERMI \cite{GLAST}.

Extrapolation of a halo model like that
of eq.~(\ref{eq:NFW}) into the Galactic center region
gives $\overline{J}_5\approx 10^3$,
large enough to produce
observable signals for many interesting choices
of $\svave$ and $m$ \cite{BUB-98}.
On the other hand, the detections by the Whipple and
H.E.S.S. collaborations of $\gamma$ rays from the Galactic center 
with energies up to 10 \tev \cite{Whipple-04,HESS-04} 
would require very large values of
$\svave\overline{J}$ \cite{hooper-04,horns-05,profumo-05},
motivating the exploration of models in which the central
dark matter density is enhanced with respect to standard
models -- for instance, via the collisionless ``spikes'' discussed above.

\begin{figure}[t]
\includegraphics[width=2.45in,angle=0.]{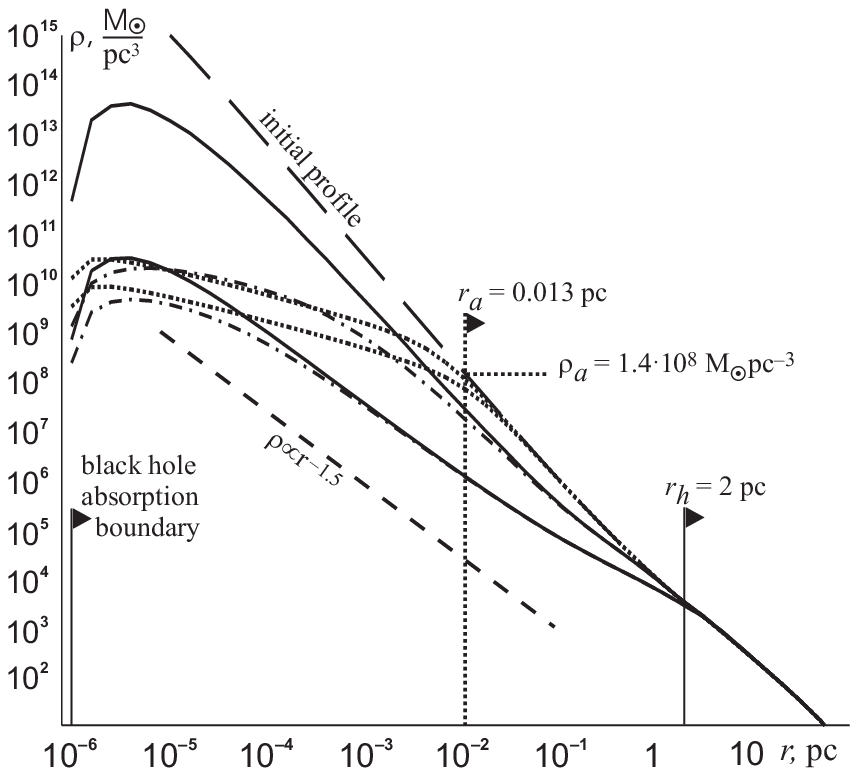}
\includegraphics[width=2.45in,angle=0.]{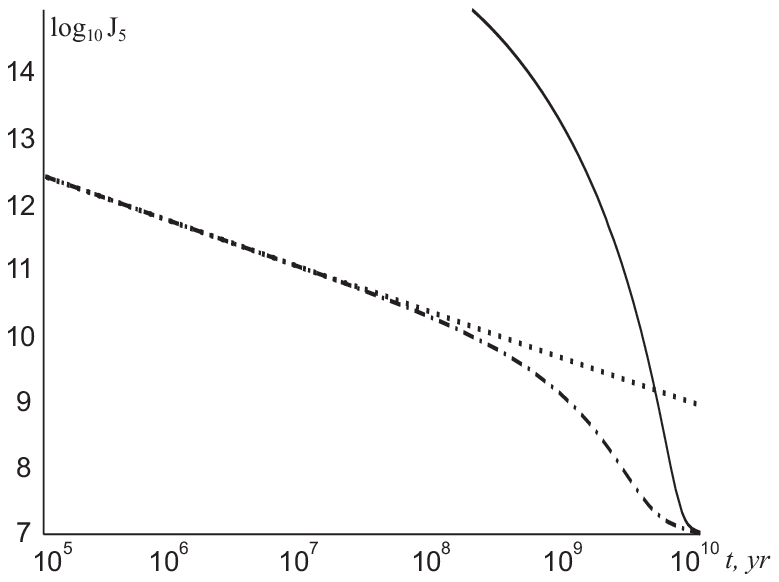}
\caption{{\it Left:} Time evolution of the dark matter density at the center
of the Milky Way, including the effects of scattering off of
stars; self-annihilations; and absorption into the SMBH.
The initial density profile, shown by the dashed line,
represents an adiabatically-compressed NFW profile with
$\rho\sim r^{-2.3}$.
The solid, dotted, and dash-dotted lines show evolved density profiles
assuming scattering only; annihilation only; and both annihilation
and scattering, repectively.
The top curve in each set is for $t=2.5$ Gyr and the bottom curve
is for $t=10$ Gyr.
{\it Right:} Evolution of the form factor 
$\overline{J}_5$ for the models in the left panel.
In  the absence of self-annihilations (solid line), 
the central density remains very high until $\sim 1$ relaxation time 
has elapsed, yielding large $J_5$ values.
When annihilations are included (dotted, dash-dotted lines),
the dark matter density near the SMBH drops rapidly and
$J_5$ is much smaller at early times.
The almost equal values of $J_5$ at $10$ Gyr in the two models
that include scattering is partly a coincidence due to the assumed
value of the annihilation cross section.
(Adapted from \cite{vasiliev-08}.)
}
\label{fig:merritt_5}
\end{figure}

In models with such high initial densities, 
$\rho_{\rm DM}$ evolves rapidly near the SMBH,
due both to scattering off of stars and to a number of
other processes.

\begin{itemize}
\item {\it Self-annihilations:} The same annihilations that
produce observable radiation also cause the dark matter density to decay.
Self-annihilations limit the density to 
$\rho_{\rm ann}\approx m/\svave t$,
% inside a region $r_{\rm ann}$,
with $t\approx 10$ Gyr the time since formation of the spike
\cite{berezinsky-92}.
The result is a weak, $\sim r^{-0.5}$ density plateau
near the SMBH \cite{vasiliev-07}.
Assuming a ``maximal'' $\svave\approx 3\times 10^{-26}$ cm$^3$ s$^{-1}$,
appropriate for a thermal relic,
and a ``minimal'' $\mchi\approx 50$ GeV, appropriate to neutralinos
in theories with gaugino and sfermion mass unification at the GUT scale 
\cite{eidelman-04}, implies $\rho_{\rm ann}\approx 10^6 M_\odot$ pc$^{-3}$
at $t=10 $ Gyr.

\item {\it Capture of dark matter within stars.}
Another potential loss term for the dark matter is 
capture {\it within} stars, due to scattering off nuclei
followed by annihilation in stellar cores.
%\cite{Chap:Bertone}.
However this process is not likely to be important
unless the cross section for WIMP-on-proton scattering is very large.

\item {\it Capture of dark matter within the SMBH.}
Any dark matter particles on orbits that intersect the 
SMBH are lost in a single orbital period.
Subsequently, scattering of dark matter particles by
stars drives a continuous flux of dark matter into 
the SMBH \cite{berezinsky-94}.
Changes in orbital angular momentum dominate the flux;
in a time $\sim t_r$, most of the dark matter within
$r_h$ will have been lost, although the net change in the 
dark matter density profile will be more modest than this
suggests 
since more particles are continuously being scattered 
onto depleted orbits \cite{merritt-04}.

\end{itemize}

\noindent
A strict inner cut-off to the dark matter density is set by the
SMBH's event horizon, $r_{Sch}=2GM_\bullet/c^2\approx 3\times 10^{-7}$ pc
although for reasonable values of
$\mchi$ and $\svave$, the density is limited by
self-annihilations well outside of $r_{Sch}$.

These various effects
can be modelled in a time-dependent way via the orbit-averaged
Fokker-Planck (FP) equation \cite{merritt-04}.
In its simplest, energy-dependent form,
the FP equation can be written
\begin{equation}
{\partial f\over\partial t} = -{1\over 4\pi^2p} 
{\partial F_E\over\partial E} - 
f(E)\nu_{\rm coll}(E) - f(E)\nu_{\rm lc}(E).
\label{eq:fp}
\end{equation}
Here $f(E)$ is the phase-space mass density of dark matter,
$E\equiv -v^2/2+\phi(r)$ is the energy per unit mass of a 
dark-matter particle, $p$ is a phase-space volume element,
and $\phi(r)$ is the gravitational
potential generated by the stars and the SMBH.
$F_E$ is the flux of particles in energy space and depends
on the stellar density profile and the stellar masses.
The two loss terms, $\nu_{\rm coll}$ and $\nu_{\rm lc}$,
represent decay of dark matter due to self-annihilations and
capture within stars;
and loss of particles into the SMBH, respectively.
A similar equation describes the evolution of the stellar
distribution \cite{MHB-07}.

Self-annihilations add a second time scale to the problem
that depends on the particle physics parameters,
the annihilation time $t_{\rm ann}$:
\begin{eqnarray}
t_{\rm ann} &\equiv& \left(\rho_{\rm DM}{\svave\over\mchi}\right)^{-1} \nonumber \\
&\approx& 0.8\ {\rm Gyr} 
\left({\mchi\over 100\ {\rm GeV}}\right)
\left({\rho_{\rm DM}\over 10^{10} M_\odot {\rm pc}^{-3}}\right)^{-1}
\left({\svave\over 10^{-26} {\rm cm}^3 {\rm s}^{-1}}\right)^{-1}.
\end{eqnarray}
Even assuming the ``maximal'' annihilation model defined above,
$t_{\rm ann}$ becomes comparable to $t_r$ only for dark matter 
densities greater than $\sim 10^8M_\odot$ pc$^{-3}$: corresponding to 
$r\lesssim 10^{-2}$ pc in models with a ``spike.''
Fig.~\ref{fig:merritt_5} illustrates this.
%Initial conditions consisted of a $\rho\sim r^{-2.33}$ ``spike'',
%produced by adiabatically compressing a $\rho\sim r^{-1}$ cusp.
At early times, annihilations dominate the changes in $\overline{J}$,
but after $\sim 1$ Gyr heating of dark matter particles by stars tends
to repopulate orbits near the SMBH, tending toward 
a $\rho_{\rm DM}\sim r^{-3/2}$ ``crest'' density profile at $r\lesssim r_h$.

The predicted spectrum of annihilation products depends
separately on $\mchi$ (shape) and $\svave\overline{J}$ (amplitude)
\cite{bertone-07};
while for a given initial dark matter model,
the final distribution of mass in the
evolutionary models depends on 
$t_r/t_{\rm ann}\propto\svave/\mchi$.
Assuming a dark matter origin for the \tev gamma rays observed
by H.E.S.S. \cite{HESS-04}, the spectrum implies
$10\tev\lesssim \mchi \lesssim 20\tev$ and
$\svave\overline{J}_5 \approx 10^{-18}$ cm$^3$ s$^{-1}$ \cite{profumo-05}.
The latter value requires either $\overline{J}_5\gsim 10^6$,
or a substantial enhancement in the dark matter relic abundance
compared with expectations for thermal freeze-out.
Fig.~\ref{fig:merritt_5} suggests that a dark matter ``crest''
can yield sufficiently high values of $\overline{J}$ if the
initial density profile is sufficiently steep.

\section {Dwarf Spheroidal Galaxies}
\label{ssec:dsph}

As noted above, dwarf spheroidal (dSph) galaxies 
depart systematically from the scaling relations obeyed 
by other ``hot'' stellar systems between size, mass and luminosity
\cite{forbes-08}.
dSphs have total luminosities and internal velocity dispersions
comparable with those of globular star clusters, but are much 
larger, implying very high ratios of (dynamical) mass to (stellar) light,
roughly $10-100$ times the Solar value \cite{mateo-98}.
Since the stars in these systems contribute a negligible fraction
of the total mass, dSphs are ideal test-beds for theories of dark matter:
in principle, $\rho_{\rm DM}(r)$ can be mapped directly given
sufficiently large samples of radial velocities \cite{merritt-93},
without the need to correct for baryon ``contamination.''

Modelling of this sort has been carried out now for roughly a dozen
dSphs \cite[e.g.][]{kleyna-02,masch-06,walker-07}.
In practice, the kinematical data are not copious enough for a
fully nonparametric approach and the inferred densities are still
somewhat model-dependent.
Interestingly, the kinematical data seem rarely if ever to demand
cusped dark matter density profiles like that of eq.~(\ref{eq:NFW}).
Halo models with low-density cores, e.g. eq.~(\ref{eq:Burkert}),
are sometimes preferred \cite{kleyna-03,goerdt-06}.
Assuming that $\rho_{\rm DM}(r)$ follows an NFW profile, 
inferred mean densities within 10 pc are $\sim 10 M_\odot$ pc$^{-3}$;
compared with $\sim 0.1 M_\odot$ pc$^{-3}$ if the inner density profile
is flat.
Inferred {\it total} masses depend less strongly on the assumed 
profile shape \cite{gilmore-07}.
%Stellar densities in these systems do not exceed 
%$\sim 1 M_\odot$ pc$^{-3}$.
Thus, while the existence of large amounts of dark matter 
is clearly established in the dSph galaxies,
as expected based on the $N$-body models,
the detailed distribution of mass within these systems
appears to be at odds with the $N$-body predictions.
dSph galaxies are similar to low-surface-brightness spiral galaxies
in this regard.

The number of known dSph satellites of the Local Group has
roughly doubled during the last decade \cite{belo-07}
and will probably continue to rise;
at last count the Milky Way halo contained at least 18 dSphs.
Their proximity, combined with their large masses,
make them good candidates for indirect detection studies
\cite{baltz-00,strigari-07},
although the predicted fluxes are interestingly large only
if the dark matter distribution is cusped \cite{evans-04}.

Angular sizes are small enough that a large fraction of their
dark matter could be imaged in a single pointing of a telescope like FERMI.
The predicted annihilation flux then scales simply as
$\sim \rho_s^2 r_s^3$ where $\rho_s$ is the dark matter density
at the scale radius $r_s$.
If the dark matter is clumped, fluxes could be boosted by up
to two orders of magnitude \cite{strigari-08}.
In addition, the low baryonic densities imply that dSphs 
should be relatively free of intrinsic gamma rays from other astrophysical
sources, making the interpretation of a signal much more straightforward
than in the case of the Galactic center.
A particularly attractive object is the recently-discovered
object Willman 1
with a luminosity of only $\sim 10^3 L_\odot$ \cite{willman-05}.
At a distance of 38 kpc,
this object is so close that it could be marginally resolved
by FERMI, in principle allowing a determination of the dark matter
distribution \cite{strigari-08}.

\bigskip

I think A. Graham and E. Vasiliev for supplying modified
versions of figures from their published work.

\bibliography{merritt}% Produces the bibliography via BibTeX.

\begin{thebibliography}{107}
\expandafter\ifx\csname natexlab\endcsname\relax\def\natexlab#1{#1}\fi
\expandafter\ifx\csname bibnamefont\endcsname\relax
  \def\bibnamefont#1{#1}\fi
\expandafter\ifx\csname bibfnamefont\endcsname\relax
  \def\bibfnamefont#1{#1}\fi
\expandafter\ifx\csname citenamefont\endcsname\relax
  \def\citenamefont#1{#1}\fi
\expandafter\ifx\csname url\endcsname\relax
  \def\url#1{\texttt{#1}}\fi
\expandafter\ifx\csname urlprefix\endcsname\relax\def\urlprefix{URL }\fi
\providecommand{\bibinfo}[2]{#2}
\providecommand{\eprint}[2][]{\url{#2}}

\bibitem[{\citenamefont{{Dubinski} and {Carlberg}}(1991)}]{DC-91}
\bibinfo{author}{\bibfnamefont{J.}~\bibnamefont{{Dubinski}}} \bibnamefont{and}
  \bibinfo{author}{\bibfnamefont{R.~G.} \bibnamefont{{Carlberg}}},
  \bibinfo{journal}{Astrophysical Journal} \textbf{\bibinfo{volume}{378}},
  \bibinfo{pages}{496} (\bibinfo{year}{1991}).

\bibitem[{\citenamefont{{Navarro} et~al.}(1996)\citenamefont{{Navarro},
  {Frenk}, and {White}}}]{NFW-96}
\bibinfo{author}{\bibfnamefont{J.~F.} \bibnamefont{{Navarro}}},
  \bibinfo{author}{\bibfnamefont{C.~S.} \bibnamefont{{Frenk}}},
  \bibnamefont{and} \bibinfo{author}{\bibfnamefont{S.~D.~M.}
  \bibnamefont{{White}}}, \bibinfo{journal}{Astrophys. J.}
  \textbf{\bibinfo{volume}{462}}, \bibinfo{pages}{563} (\bibinfo{year}{1996}),
  \eprint{arXiv:astro-ph/9508025}.

\bibitem[{\citenamefont{{Ghigna} et~al.}(1998)\citenamefont{{Ghigna}, {Moore},
  {Governato}, {Lake}, {Quinn}, and {Stadel}}}]{ghigna-98}
\bibinfo{author}{\bibfnamefont{S.}~\bibnamefont{{Ghigna}}},
  \bibinfo{author}{\bibfnamefont{B.}~\bibnamefont{{Moore}}},
  \bibinfo{author}{\bibfnamefont{F.}~\bibnamefont{{Governato}}},
  \bibinfo{author}{\bibfnamefont{G.}~\bibnamefont{{Lake}}},
  \bibinfo{author}{\bibfnamefont{T.}~\bibnamefont{{Quinn}}}, \bibnamefont{and}
  \bibinfo{author}{\bibfnamefont{J.}~\bibnamefont{{Stadel}}},
  \bibinfo{journal}{Mon. Not. R. Astron. Soc.} \textbf{\bibinfo{volume}{300}},
  \bibinfo{pages}{146} (\bibinfo{year}{1998}), \eprint{arXiv:astro-ph/9801192}.

\bibitem[{\citenamefont{{Moore} et~al.}(1998)\citenamefont{{Moore},
  {Governato}, {Quinn}, {Stadel}, and {Lake}}}]{moore-98}
\bibinfo{author}{\bibfnamefont{B.}~\bibnamefont{{Moore}}},
  \bibinfo{author}{\bibfnamefont{F.}~\bibnamefont{{Governato}}},
  \bibinfo{author}{\bibfnamefont{T.}~\bibnamefont{{Quinn}}},
  \bibinfo{author}{\bibfnamefont{J.}~\bibnamefont{{Stadel}}}, \bibnamefont{and}
  \bibinfo{author}{\bibfnamefont{G.}~\bibnamefont{{Lake}}},
  \bibinfo{journal}{Astrophysical Journal Letters}
  \textbf{\bibinfo{volume}{499}}, \bibinfo{pages}{L5+} (\bibinfo{year}{1998}),
  \eprint{arXiv:astro-ph/9709051}.

\bibitem[{\citenamefont{{Klypin} et~al.}(2001)\citenamefont{{Klypin},
  {Kravtsov}, {Bullock}, and {Primack}}}]{klypin-01}
\bibinfo{author}{\bibfnamefont{A.}~\bibnamefont{{Klypin}}},
  \bibinfo{author}{\bibfnamefont{A.~V.} \bibnamefont{{Kravtsov}}},
  \bibinfo{author}{\bibfnamefont{J.~S.} \bibnamefont{{Bullock}}},
  \bibnamefont{and} \bibinfo{author}{\bibfnamefont{J.~R.}
  \bibnamefont{{Primack}}}, \bibinfo{journal}{Astrophysical Journal}
  \textbf{\bibinfo{volume}{554}}, \bibinfo{pages}{903} (\bibinfo{year}{2001}),
  \eprint{arXiv:astro-ph/0006343}.

\bibitem[{\citenamefont{{Diemand} et~al.}(2004)\citenamefont{{Diemand},
  {Moore}, and {Stadel}}}]{diemand-04}
\bibinfo{author}{\bibfnamefont{J.}~\bibnamefont{{Diemand}}},
  \bibinfo{author}{\bibfnamefont{B.}~\bibnamefont{{Moore}}}, \bibnamefont{and}
  \bibinfo{author}{\bibfnamefont{J.}~\bibnamefont{{Stadel}}},
  \bibinfo{journal}{Mon. Not. R. Astron. Soc.} \textbf{\bibinfo{volume}{353}},
  \bibinfo{pages}{624} (\bibinfo{year}{2004}), \eprint{arXiv:astro-ph/0402267}.

\bibitem[{\citenamefont{{Navarro} et~al.}(2004)\citenamefont{{Navarro},
  {Hayashi}, {Power}, {Jenkins}, {Frenk}, {White}, {Springel}, {Stadel}, and
  {Quinn}}}]{navarro-04}
\bibinfo{author}{\bibfnamefont{J.~F.} \bibnamefont{{Navarro}}},
  \bibinfo{author}{\bibfnamefont{E.}~\bibnamefont{{Hayashi}}},
  \bibinfo{author}{\bibfnamefont{C.}~\bibnamefont{{Power}}},
  \bibinfo{author}{\bibfnamefont{A.~R.} \bibnamefont{{Jenkins}}},
  \bibinfo{author}{\bibfnamefont{C.~S.} \bibnamefont{{Frenk}}},
  \bibinfo{author}{\bibfnamefont{S.~D.~M.} \bibnamefont{{White}}},
  \bibinfo{author}{\bibfnamefont{V.}~\bibnamefont{{Springel}}},
  \bibinfo{author}{\bibfnamefont{J.}~\bibnamefont{{Stadel}}}, \bibnamefont{and}
  \bibinfo{author}{\bibfnamefont{T.~R.} \bibnamefont{{Quinn}}},
  \bibinfo{journal}{Mon. Not. R. Astron. Soc.} \textbf{\bibinfo{volume}{349}},
  \bibinfo{pages}{1039} (\bibinfo{year}{2004}).

\bibitem[{\citenamefont{{Merritt} et~al.}(2005)\citenamefont{{Merritt},
  {Navarro}, {Ludlow}, and {Jenkins}}}]{merritt-05}
\bibinfo{author}{\bibfnamefont{D.}~\bibnamefont{{Merritt}}},
  \bibinfo{author}{\bibfnamefont{J.~F.} \bibnamefont{{Navarro}}},
  \bibinfo{author}{\bibfnamefont{A.}~\bibnamefont{{Ludlow}}}, \bibnamefont{and}
  \bibinfo{author}{\bibfnamefont{A.}~\bibnamefont{{Jenkins}}},
  \bibinfo{journal}{Astrophys. J. (Lett.)} \textbf{\bibinfo{volume}{624}},
  \bibinfo{pages}{L85} (\bibinfo{year}{2005}).

\bibitem[{\citenamefont{{Navarro} et~al.}(2008)\citenamefont{{Navarro},
  {Ludlow}, {Springel}, {Wang}, {Vogelsberger}, {White}, {Jenkins}, {Frenk},
  and {Helmi}}}]{navarro-08}
\bibinfo{author}{\bibfnamefont{J.~F.} \bibnamefont{{Navarro}}},
  \bibinfo{author}{\bibfnamefont{A.}~\bibnamefont{{Ludlow}}},
  \bibinfo{author}{\bibfnamefont{V.}~\bibnamefont{{Springel}}},
  \bibinfo{author}{\bibfnamefont{J.}~\bibnamefont{{Wang}}},
  \bibinfo{author}{\bibfnamefont{M.}~\bibnamefont{{Vogelsberger}}},
  \bibinfo{author}{\bibfnamefont{S.~D.~M.} \bibnamefont{{White}}},
  \bibinfo{author}{\bibfnamefont{A.}~\bibnamefont{{Jenkins}}},
  \bibinfo{author}{\bibfnamefont{C.~S.} \bibnamefont{{Frenk}}},
  \bibnamefont{and} \bibinfo{author}{\bibfnamefont{A.}~\bibnamefont{{Helmi}}},
  \bibinfo{journal}{ArXiv e-prints}  (\bibinfo{year}{2008}),
  \eprint{0810.1522}.

\bibitem[{\citenamefont{{S\'ersic}}(1968)}]{sersic-68}
\bibinfo{author}{\bibfnamefont{J.~L.} \bibnamefont{{S\'ersic}}},
  \emph{\bibinfo{title}{{Atlas de galaxias australes}}}
  (\bibinfo{publisher}{Cordoba, Argentina: Observatorio Astronomico, 1968},
  \bibinfo{year}{1968}).

\bibitem[{\citenamefont{{de Vaucouleurs}}(1948)}]{devauc-48}
\bibinfo{author}{\bibfnamefont{G.}~\bibnamefont{{de Vaucouleurs}}},
  \bibinfo{journal}{Annales d'Astrophysique} \textbf{\bibinfo{volume}{11}},
  \bibinfo{pages}{247} (\bibinfo{year}{1948}).

\bibitem[{\citenamefont{{B{\"o}ker} et~al.}(2002)\citenamefont{{B{\"o}ker},
  {Laine}, {van der Marel}, {Sarzi}, {Rix}, {Ho}, and {Shields}}}]{boeker-02}
\bibinfo{author}{\bibfnamefont{T.}~\bibnamefont{{B{\"o}ker}}},
  \bibinfo{author}{\bibfnamefont{S.}~\bibnamefont{{Laine}}},
  \bibinfo{author}{\bibfnamefont{R.~P.} \bibnamefont{{van der Marel}}},
  \bibinfo{author}{\bibfnamefont{M.}~\bibnamefont{{Sarzi}}},
  \bibinfo{author}{\bibfnamefont{H.-W.} \bibnamefont{{Rix}}},
  \bibinfo{author}{\bibfnamefont{L.~C.} \bibnamefont{{Ho}}}, \bibnamefont{and}
  \bibinfo{author}{\bibfnamefont{J.~C.} \bibnamefont{{Shields}}},
  \bibinfo{journal}{Astron. J.} \textbf{\bibinfo{volume}{123}},
  \bibinfo{pages}{1389} (\bibinfo{year}{2002}),
  \eprint{arXiv:astro-ph/0112086}.

\bibitem[{\citenamefont{{C{\^o}t{\'e}}
  et~al.}(2006)\citenamefont{{C{\^o}t{\'e}}, {Piatek}, {Ferrarese},
  {Jord{\'a}n}, {Merritt}, {Peng}, {Ha{\c s}egan}, {Blakeslee}, {Mei}, {West}
  et~al.}}]{ACS8}
\bibinfo{author}{\bibfnamefont{P.}~\bibnamefont{{C{\^o}t{\'e}}}},
  \bibinfo{author}{\bibfnamefont{S.}~\bibnamefont{{Piatek}}},
  \bibinfo{author}{\bibfnamefont{L.}~\bibnamefont{{Ferrarese}}},
  \bibinfo{author}{\bibfnamefont{A.}~\bibnamefont{{Jord{\'a}n}}},
  \bibinfo{author}{\bibfnamefont{D.}~\bibnamefont{{Merritt}}},
  \bibinfo{author}{\bibfnamefont{E.~W.} \bibnamefont{{Peng}}},
  \bibinfo{author}{\bibfnamefont{M.}~\bibnamefont{{Ha{\c s}egan}}},
  \bibinfo{author}{\bibfnamefont{J.~P.} \bibnamefont{{Blakeslee}}},
  \bibinfo{author}{\bibfnamefont{S.}~\bibnamefont{{Mei}}},
  \bibinfo{author}{\bibfnamefont{M.~J.} \bibnamefont{{West}}},
  \bibnamefont{et~al.}, \bibinfo{journal}{Astrophys. J. Suppl.}
  \textbf{\bibinfo{volume}{165}}, \bibinfo{pages}{57} (\bibinfo{year}{2006}),
  \eprint{arXiv:astro-ph/0603252}.

\bibitem[{\citenamefont{{C{\^o}t{\'e}}
  et~al.}(2007)\citenamefont{{C{\^o}t{\'e}}, {Ferrarese}, {Jord{\'a}n},
  {Blakeslee}, {Chen}, {Infante}, {Merritt}, {Mei}, {Peng}, {Tonry}
  et~al.}}]{ACSFornax2}
\bibinfo{author}{\bibfnamefont{P.}~\bibnamefont{{C{\^o}t{\'e}}}},
  \bibinfo{author}{\bibfnamefont{L.}~\bibnamefont{{Ferrarese}}},
  \bibinfo{author}{\bibfnamefont{A.}~\bibnamefont{{Jord{\'a}n}}},
  \bibinfo{author}{\bibfnamefont{J.~P.} \bibnamefont{{Blakeslee}}},
  \bibinfo{author}{\bibfnamefont{C.-W.} \bibnamefont{{Chen}}},
  \bibinfo{author}{\bibfnamefont{L.}~\bibnamefont{{Infante}}},
  \bibinfo{author}{\bibfnamefont{D.}~\bibnamefont{{Merritt}}},
  \bibinfo{author}{\bibfnamefont{S.}~\bibnamefont{{Mei}}},
  \bibinfo{author}{\bibfnamefont{E.~W.} \bibnamefont{{Peng}}},
  \bibinfo{author}{\bibfnamefont{J.~L.} \bibnamefont{{Tonry}}},
  \bibnamefont{et~al.}, \bibinfo{journal}{Astrophys. J.}
  \textbf{\bibinfo{volume}{671}}, \bibinfo{pages}{1456} (\bibinfo{year}{2007}),
  \eprint{0711.1358}.

\bibitem[{\citenamefont{{Sch{\"o}del}
  et~al.}(2008{\natexlab{a}})\citenamefont{{Sch{\"o}del}, {Merritt}, and
  {Eckart}}}]{schoedel-08}
\bibinfo{author}{\bibfnamefont{R.}~\bibnamefont{{Sch{\"o}del}}},
  \bibinfo{author}{\bibfnamefont{D.}~\bibnamefont{{Merritt}}},
  \bibnamefont{and} \bibinfo{author}{\bibfnamefont{A.}~\bibnamefont{{Eckart}}},
  \bibinfo{journal}{Journal of Physics Conference Series}
  \textbf{\bibinfo{volume}{131}}, \bibinfo{pages}{012044}
  (\bibinfo{year}{2008}{\natexlab{a}}), \eprint{0810.0204}.

\bibitem[{\citenamefont{{Merritt} and {Ferrarese}}(2001)}]{MF-01c}
\bibinfo{author}{\bibfnamefont{D.}~\bibnamefont{{Merritt}}} \bibnamefont{and}
  \bibinfo{author}{\bibfnamefont{L.}~\bibnamefont{{Ferrarese}}}, in
  \emph{\bibinfo{booktitle}{ASP Conf. Ser. 249: The Central Kiloparsec of
  Starbursts and AGN: The La Palma Connection}} (\bibinfo{year}{2001}), pp.
  \bibinfo{pages}{335--+}.

\bibitem[{\citenamefont{{Ferrarese} and {Ford}}(2005)}]{FF-05}
\bibinfo{author}{\bibfnamefont{L.}~\bibnamefont{{Ferrarese}}} \bibnamefont{and}
  \bibinfo{author}{\bibfnamefont{H.}~\bibnamefont{{Ford}}},
  \bibinfo{journal}{Space Science Reviews} \textbf{\bibinfo{volume}{116}},
  \bibinfo{pages}{523} (\bibinfo{year}{2005}).

\bibitem[{\citenamefont{{Gillessen} et~al.}(2008)\citenamefont{{Gillessen},
  {Eisenhauer}, {Trippe}, {Alexander}, {Genzel}, {Martins}, and
  {Ott}}}]{gillessen-08}
\bibinfo{author}{\bibfnamefont{S.}~\bibnamefont{{Gillessen}}},
  \bibinfo{author}{\bibfnamefont{F.}~\bibnamefont{{Eisenhauer}}},
  \bibinfo{author}{\bibfnamefont{S.}~\bibnamefont{{Trippe}}},
  \bibinfo{author}{\bibfnamefont{T.}~\bibnamefont{{Alexander}}},
  \bibinfo{author}{\bibfnamefont{R.}~\bibnamefont{{Genzel}}},
  \bibinfo{author}{\bibfnamefont{F.}~\bibnamefont{{Martins}}},
  \bibnamefont{and} \bibinfo{author}{\bibfnamefont{T.}~\bibnamefont{{Ott}}},
  \bibinfo{journal}{ArXiv e-prints}  (\bibinfo{year}{2008}),
  \eprint{0810.4674}.

\bibitem[{\citenamefont{{Ghez} et~al.}(2008)\citenamefont{{Ghez}, {Salim},
  {Weinberg}, {Lu}, {Do}, {Dunn}, {Matthews}, {Morris}, {Yelda}, {Becklin}
  et~al.}}]{ghez-08}
\bibinfo{author}{\bibfnamefont{A.~M.} \bibnamefont{{Ghez}}},
  \bibinfo{author}{\bibfnamefont{S.}~\bibnamefont{{Salim}}},
  \bibinfo{author}{\bibfnamefont{N.~N.} \bibnamefont{{Weinberg}}},
  \bibinfo{author}{\bibfnamefont{J.~R.} \bibnamefont{{Lu}}},
  \bibinfo{author}{\bibfnamefont{T.}~\bibnamefont{{Do}}},
  \bibinfo{author}{\bibfnamefont{J.~K.} \bibnamefont{{Dunn}}},
  \bibinfo{author}{\bibfnamefont{K.}~\bibnamefont{{Matthews}}},
  \bibinfo{author}{\bibfnamefont{M.}~\bibnamefont{{Morris}}},
  \bibinfo{author}{\bibfnamefont{S.}~\bibnamefont{{Yelda}}},
  \bibinfo{author}{\bibfnamefont{E.~E.} \bibnamefont{{Becklin}}},
  \bibnamefont{et~al.}, \bibinfo{journal}{ArXiv e-prints}
  (\bibinfo{year}{2008}), \eprint{0808.2870}.

\bibitem[{\citenamefont{{Miller} and {Colbert}}(2004)}]{MC-04}
\bibinfo{author}{\bibfnamefont{M.~C.} \bibnamefont{{Miller}}} \bibnamefont{and}
  \bibinfo{author}{\bibfnamefont{E.~J.~M.} \bibnamefont{{Colbert}}},
  \bibinfo{journal}{International Journal of Modern Physics D}
  \textbf{\bibinfo{volume}{13}}, \bibinfo{pages}{1} (\bibinfo{year}{2004}).

\bibitem[{\citenamefont{{Merritt}}(2006)}]{merritt-06}
\bibinfo{author}{\bibfnamefont{D.}~\bibnamefont{{Merritt}}},
  \bibinfo{journal}{The Astrophysical Journal} \textbf{\bibinfo{volume}{648}},
  \bibinfo{pages}{976} (\bibinfo{year}{2006}).

\bibitem[{\citenamefont{{Seth} et~al.}(2008)\citenamefont{{Seth},
  {Ag{\"u}eros}, {Lee}, and {Basu-Zych}}}]{seth-08}
\bibinfo{author}{\bibfnamefont{A.}~\bibnamefont{{Seth}}},
  \bibinfo{author}{\bibfnamefont{M.}~\bibnamefont{{Ag{\"u}eros}}},
  \bibinfo{author}{\bibfnamefont{D.}~\bibnamefont{{Lee}}}, \bibnamefont{and}
  \bibinfo{author}{\bibfnamefont{A.}~\bibnamefont{{Basu-Zych}}},
  \bibinfo{journal}{Astrophys. J.} \textbf{\bibinfo{volume}{678}},
  \bibinfo{pages}{116} (\bibinfo{year}{2008}), \eprint{0801.0439}.

\bibitem[{\citenamefont{{Spitzer}}(1987)}]{spitzer-87}
\bibinfo{author}{\bibfnamefont{L.}~\bibnamefont{{Spitzer}}},
  \emph{\bibinfo{title}{{Dynamical evolution of globular clusters}}}
  (\bibinfo{publisher}{Princeton, NJ, Princeton University Press, 1987, 191
  p.}, \bibinfo{year}{1987}).

\bibitem[{\citenamefont{{Merritt}
  et~al.}(2007{\natexlab{a}})\citenamefont{{Merritt}, {Mikkola}, and
  {Szell}}}]{MMS-07}
\bibinfo{author}{\bibfnamefont{D.}~\bibnamefont{{Merritt}}},
  \bibinfo{author}{\bibfnamefont{S.}~\bibnamefont{{Mikkola}}},
  \bibnamefont{and} \bibinfo{author}{\bibfnamefont{A.}~\bibnamefont{{Szell}}},
  \bibinfo{journal}{Astrophys. J.} \textbf{\bibinfo{volume}{671}},
  \bibinfo{pages}{53} (\bibinfo{year}{2007}{\natexlab{a}}), \eprint{0705.2745}.

\bibitem[{\citenamefont{{Merritt}}(2008)}]{merritt-09}
\bibinfo{author}{\bibfnamefont{D.}~\bibnamefont{{Merritt}}},
  \bibinfo{journal}{ArXiv e-prints}  (\bibinfo{year}{2008}),
  \eprint{0802.3186}.

\bibitem[{\citenamefont{{Sch{\"o}del} et~al.}(2007)\citenamefont{{Sch{\"o}del},
  {Eckart}, {Alexander}, {Merritt}, {Genzel}, {Sternberg}, {Meyer}, {Kul},
  {Moultaka}, {Ott} et~al.}}]{schoedel-07}
\bibinfo{author}{\bibfnamefont{R.}~\bibnamefont{{Sch{\"o}del}}},
  \bibinfo{author}{\bibfnamefont{A.}~\bibnamefont{{Eckart}}},
  \bibinfo{author}{\bibfnamefont{T.}~\bibnamefont{{Alexander}}},
  \bibinfo{author}{\bibfnamefont{D.}~\bibnamefont{{Merritt}}},
  \bibinfo{author}{\bibfnamefont{R.}~\bibnamefont{{Genzel}}},
  \bibinfo{author}{\bibfnamefont{A.}~\bibnamefont{{Sternberg}}},
  \bibinfo{author}{\bibfnamefont{L.}~\bibnamefont{{Meyer}}},
  \bibinfo{author}{\bibfnamefont{F.}~\bibnamefont{{Kul}}},
  \bibinfo{author}{\bibfnamefont{J.}~\bibnamefont{{Moultaka}}},
  \bibinfo{author}{\bibfnamefont{T.}~\bibnamefont{{Ott}}},
  \bibnamefont{et~al.}, \bibinfo{journal}{Astron. Ap.}
  \textbf{\bibinfo{volume}{469}}, \bibinfo{pages}{125} (\bibinfo{year}{2007}),
  \eprint{arXiv:astro-ph/0703178}.

\bibitem[{\citenamefont{{Bahcall} and {Wolf}}(1976)}]{BW-76}
\bibinfo{author}{\bibfnamefont{J.~N.} \bibnamefont{{Bahcall}}}
  \bibnamefont{and} \bibinfo{author}{\bibfnamefont{R.~A.}
  \bibnamefont{{Wolf}}}, \bibinfo{journal}{Astrophys. J.}
  \textbf{\bibinfo{volume}{209}}, \bibinfo{pages}{214} (\bibinfo{year}{1976}).

\bibitem[{\citenamefont{{Bahcall} and {Wolf}}(1977)}]{BW-77}
\bibinfo{author}{\bibfnamefont{J.~N.} \bibnamefont{{Bahcall}}}
  \bibnamefont{and} \bibinfo{author}{\bibfnamefont{R.~A.}
  \bibnamefont{{Wolf}}}, \bibinfo{journal}{Astrophys. J.}
  \textbf{\bibinfo{volume}{216}}, \bibinfo{pages}{883} (\bibinfo{year}{1977}).

\bibitem[{\citenamefont{{Graham} et~al.}(2006)\citenamefont{{Graham},
  {Merritt}, {Moore}, {Diemand}, and {Terzi{\'c}}}}]{empirical3}
\bibinfo{author}{\bibfnamefont{A.~W.} \bibnamefont{{Graham}}},
  \bibinfo{author}{\bibfnamefont{D.}~\bibnamefont{{Merritt}}},
  \bibinfo{author}{\bibfnamefont{B.}~\bibnamefont{{Moore}}},
  \bibinfo{author}{\bibfnamefont{J.}~\bibnamefont{{Diemand}}},
  \bibnamefont{and}
  \bibinfo{author}{\bibfnamefont{B.}~\bibnamefont{{Terzi{\'c}}}},
  \bibinfo{journal}{Astron. J.} \textbf{\bibinfo{volume}{132}},
  \bibinfo{pages}{2711} (\bibinfo{year}{2006}),
  \eprint{arXiv:astro-ph/0608614}.

\bibitem[{\citenamefont{{Burkert}}(1995)}]{Burkert-95}
\bibinfo{author}{\bibfnamefont{A.}~\bibnamefont{{Burkert}}},
  \bibinfo{journal}{ApJ Letters} \textbf{\bibinfo{volume}{447}},
  \bibinfo{pages}{L25+} (\bibinfo{year}{1995}).

\bibitem[{\citenamefont{{Salucci} and {Burkert}}(2000)}]{SB-00}
\bibinfo{author}{\bibfnamefont{P.}~\bibnamefont{{Salucci}}} \bibnamefont{and}
  \bibinfo{author}{\bibfnamefont{A.}~\bibnamefont{{Burkert}}},
  \bibinfo{journal}{ApJ Letters} \textbf{\bibinfo{volume}{537}},
  \bibinfo{pages}{L9} (\bibinfo{year}{2000}).

\bibitem[{\citenamefont{{de Blok} and {Bosma}}(2002)}]{Blok-02}
\bibinfo{author}{\bibfnamefont{W.~J.~G.} \bibnamefont{{de Blok}}}
  \bibnamefont{and} \bibinfo{author}{\bibfnamefont{A.}~\bibnamefont{{Bosma}}},
  \bibinfo{journal}{Astronomy and Astrophysics} \textbf{\bibinfo{volume}{385}},
  \bibinfo{pages}{816} (\bibinfo{year}{2002}).

\bibitem[{\citenamefont{{Gentile} et~al.}(2005)\citenamefont{{Gentile},
  {Burkert}, {Salucci}, {Klein}, and {Walter}}}]{Gentile-05}
\bibinfo{author}{\bibfnamefont{G.}~\bibnamefont{{Gentile}}},
  \bibinfo{author}{\bibfnamefont{A.}~\bibnamefont{{Burkert}}},
  \bibinfo{author}{\bibfnamefont{P.}~\bibnamefont{{Salucci}}},
  \bibinfo{author}{\bibfnamefont{U.}~\bibnamefont{{Klein}}}, \bibnamefont{and}
  \bibinfo{author}{\bibfnamefont{F.}~\bibnamefont{{Walter}}},
  \bibinfo{journal}{ApJ Letters} \textbf{\bibinfo{volume}{634}},
  \bibinfo{pages}{L145} (\bibinfo{year}{2005}).

\bibitem[{\citenamefont{{de Blok}}(2005)}]{Blok-05}
\bibinfo{author}{\bibfnamefont{W.~J.~G.} \bibnamefont{{de Blok}}},
  \bibinfo{journal}{Astrophys. J.} \textbf{\bibinfo{volume}{634}},
  \bibinfo{pages}{227} (\bibinfo{year}{2005}).

\bibitem[{\citenamefont{{Spekkens} et~al.}(2005)\citenamefont{{Spekkens},
  {Giovanelli}, and {Haynes}}}]{SGH-05}
\bibinfo{author}{\bibfnamefont{K.}~\bibnamefont{{Spekkens}}},
  \bibinfo{author}{\bibfnamefont{R.}~\bibnamefont{{Giovanelli}}},
  \bibnamefont{and} \bibinfo{author}{\bibfnamefont{M.~P.}
  \bibnamefont{{Haynes}}}, \bibinfo{journal}{Astronomical Journal}
  \textbf{\bibinfo{volume}{129}}, \bibinfo{pages}{2119} (\bibinfo{year}{2005}).

\bibitem[{\citenamefont{{Simon} et~al.}(2005)\citenamefont{{Simon}, {Bolatto},
  {Leroy}, {Blitz}, and {Gates}}}]{Simon-05}
\bibinfo{author}{\bibfnamefont{J.~D.} \bibnamefont{{Simon}}},
  \bibinfo{author}{\bibfnamefont{A.~D.} \bibnamefont{{Bolatto}}},
  \bibinfo{author}{\bibfnamefont{A.}~\bibnamefont{{Leroy}}},
  \bibinfo{author}{\bibfnamefont{L.}~\bibnamefont{{Blitz}}}, \bibnamefont{and}
  \bibinfo{author}{\bibfnamefont{E.~L.} \bibnamefont{{Gates}}},
  \bibinfo{journal}{Astrophys. J.} \textbf{\bibinfo{volume}{621}},
  \bibinfo{pages}{757} (\bibinfo{year}{2005}).

\bibitem[{\citenamefont{{Valenzuela} et~al.}(2007)\citenamefont{{Valenzuela},
  {Rhee}, {Klypin}, {Governato}, {Stinson}, {Quinn}, and
  {Wadsley}}}]{valenzuela-07}
\bibinfo{author}{\bibfnamefont{O.}~\bibnamefont{{Valenzuela}}},
  \bibinfo{author}{\bibfnamefont{G.}~\bibnamefont{{Rhee}}},
  \bibinfo{author}{\bibfnamefont{A.}~\bibnamefont{{Klypin}}},
  \bibinfo{author}{\bibfnamefont{F.}~\bibnamefont{{Governato}}},
  \bibinfo{author}{\bibfnamefont{G.}~\bibnamefont{{Stinson}}},
  \bibinfo{author}{\bibfnamefont{T.}~\bibnamefont{{Quinn}}}, \bibnamefont{and}
  \bibinfo{author}{\bibfnamefont{J.}~\bibnamefont{{Wadsley}}},
  \bibinfo{journal}{Astrophys. J.} \textbf{\bibinfo{volume}{657}},
  \bibinfo{pages}{773} (\bibinfo{year}{2007}), \eprint{arXiv:astro-ph/0509644}.

\bibitem[{\citenamefont{{de Blok}}(2004)}]{Blok-04}
\bibinfo{author}{\bibfnamefont{W.~J.~G.} \bibnamefont{{de Blok}}}, in
  \emph{\bibinfo{booktitle}{IAU Symposium}}, edited by
  \bibinfo{editor}{\bibfnamefont{S.}~\bibnamefont{{Ryder}}},
  \bibinfo{editor}{\bibfnamefont{D.}~\bibnamefont{{Pisano}}},
  \bibinfo{editor}{\bibfnamefont{M.}~\bibnamefont{{Walker}}}, \bibnamefont{and}
  \bibinfo{editor}{\bibfnamefont{K.}~\bibnamefont{{Freeman}}}
  (\bibinfo{year}{2004}), pp. \bibinfo{pages}{69--+}.

\bibitem[{\citenamefont{{Einasto}}(1965)}]{einasto-65}
\bibinfo{author}{\bibfnamefont{J.}~\bibnamefont{{Einasto}}},
  \bibinfo{journal}{Trudy Inst. Astrofiz. Alma-Ata}
  \textbf{\bibinfo{volume}{5}}, \bibinfo{pages}{87} (\bibinfo{year}{1965}).

\bibitem[{\citenamefont{{Prada} et~al.}(2006)\citenamefont{{Prada}, {Klypin},
  {Simonneau}, {Betancort-Rijo}, {Patiri}, {Gottl{\"o}ber}, and
  {Sanchez-Conde}}}]{prada-06}
\bibinfo{author}{\bibfnamefont{F.}~\bibnamefont{{Prada}}},
  \bibinfo{author}{\bibfnamefont{A.~A.} \bibnamefont{{Klypin}}},
  \bibinfo{author}{\bibfnamefont{E.}~\bibnamefont{{Simonneau}}},
  \bibinfo{author}{\bibfnamefont{J.}~\bibnamefont{{Betancort-Rijo}}},
  \bibinfo{author}{\bibfnamefont{S.}~\bibnamefont{{Patiri}}},
  \bibinfo{author}{\bibfnamefont{S.}~\bibnamefont{{Gottl{\"o}ber}}},
  \bibnamefont{and} \bibinfo{author}{\bibfnamefont{M.~A.}
  \bibnamefont{{Sanchez-Conde}}}, \bibinfo{journal}{Astrophys. J.}
  \textbf{\bibinfo{volume}{645}}, \bibinfo{pages}{1001} (\bibinfo{year}{2006}),
  \eprint{arXiv:astro-ph/0506432}.

\bibitem[{\citenamefont{{Silk}}(2002)}]{silk-02}
\bibinfo{author}{\bibfnamefont{J.}~\bibnamefont{{Silk}}},
  \bibinfo{journal}{International Journal of Modern Physics A}
  \textbf{\bibinfo{volume}{17}}, \bibinfo{pages}{167} (\bibinfo{year}{2002}),
  \eprint{arXiv:astro-ph/0110404}.

\bibitem[{\citenamefont{{Tasitsiomi}}(2003)}]{tasi-03}
\bibinfo{author}{\bibfnamefont{A.}~\bibnamefont{{Tasitsiomi}}},
  \bibinfo{journal}{International Journal of Modern Physics D}
  \textbf{\bibinfo{volume}{12}}, \bibinfo{pages}{1157} (\bibinfo{year}{2003}).

\bibitem[{\citenamefont{{Sanders} and {McGaugh}}(2002)}]{SM-02}
\bibinfo{author}{\bibfnamefont{R.~H.} \bibnamefont{{Sanders}}}
  \bibnamefont{and} \bibinfo{author}{\bibfnamefont{S.~S.}
  \bibnamefont{{McGaugh}}}, \bibinfo{journal}{Ann. Rev. Astron. Astrophys.}
  \textbf{\bibinfo{volume}{40}}, \bibinfo{pages}{263} (\bibinfo{year}{2002}),
  \eprint{arXiv:astro-ph/0204521}.

\bibitem[{\citenamefont{{Blumenthal} et~al.}(1986)\citenamefont{{Blumenthal},
  {Faber}, {Flores}, and {Primack}}}]{blumenthal-86}
\bibinfo{author}{\bibfnamefont{G.~R.} \bibnamefont{{Blumenthal}}},
  \bibinfo{author}{\bibfnamefont{S.~M.} \bibnamefont{{Faber}}},
  \bibinfo{author}{\bibfnamefont{R.}~\bibnamefont{{Flores}}}, \bibnamefont{and}
  \bibinfo{author}{\bibfnamefont{J.~R.} \bibnamefont{{Primack}}},
  \bibinfo{journal}{Astrophys. J.} \textbf{\bibinfo{volume}{301}},
  \bibinfo{pages}{27} (\bibinfo{year}{1986}).

\bibitem[{\citenamefont{{Prada} et~al.}(2004)\citenamefont{{Prada}, {Klypin},
  {Flix}, {Mart{\'{\i}}nez}, and {Simonneau}}}]{prada-04}
\bibinfo{author}{\bibfnamefont{F.}~\bibnamefont{{Prada}}},
  \bibinfo{author}{\bibfnamefont{A.}~\bibnamefont{{Klypin}}},
  \bibinfo{author}{\bibfnamefont{J.}~\bibnamefont{{Flix}}},
  \bibinfo{author}{\bibfnamefont{M.}~\bibnamefont{{Mart{\'{\i}}nez}}},
  \bibnamefont{and}
  \bibinfo{author}{\bibfnamefont{E.}~\bibnamefont{{Simonneau}}},
  \bibinfo{journal}{Physical Review Letters} \textbf{\bibinfo{volume}{93}},
  \bibinfo{pages}{241301} (\bibinfo{year}{2004}).

\bibitem[{\citenamefont{{Gnedin} et~al.}(2004)\citenamefont{{Gnedin},
  {Kravtsov}, {Klypin}, and {Nagai}}}]{gnedin-04}
\bibinfo{author}{\bibfnamefont{O.~Y.} \bibnamefont{{Gnedin}}},
  \bibinfo{author}{\bibfnamefont{A.~V.} \bibnamefont{{Kravtsov}}},
  \bibinfo{author}{\bibfnamefont{A.~A.} \bibnamefont{{Klypin}}},
  \bibnamefont{and} \bibinfo{author}{\bibfnamefont{D.}~\bibnamefont{{Nagai}}},
  \bibinfo{journal}{Astrophys. J.} \textbf{\bibinfo{volume}{616}},
  \bibinfo{pages}{16} (\bibinfo{year}{2004}), \eprint{arXiv:astro-ph/0406247}.

\bibitem[{\citenamefont{{Mambrini} et~al.}(2006)\citenamefont{{Mambrini},
  {Mu{\~n}oz}, {Nezri}, and {Prada}}}]{mambrini-06}
\bibinfo{author}{\bibfnamefont{Y.}~\bibnamefont{{Mambrini}}},
  \bibinfo{author}{\bibfnamefont{C.}~\bibnamefont{{Mu{\~n}oz}}},
  \bibinfo{author}{\bibfnamefont{E.}~\bibnamefont{{Nezri}}}, \bibnamefont{and}
  \bibinfo{author}{\bibfnamefont{F.}~\bibnamefont{{Prada}}},
  \bibinfo{journal}{Journal of Cosmology and Astro-Particle Physics}
  \textbf{\bibinfo{volume}{1}}, \bibinfo{pages}{10} (\bibinfo{year}{2006}),
  \eprint{arXiv:hep-ph/0506204}.

\bibitem[{\citenamefont{{Bertone} and {Merritt}}(2005)}]{BM-05}
\bibinfo{author}{\bibfnamefont{G.}~\bibnamefont{{Bertone}}} \bibnamefont{and}
  \bibinfo{author}{\bibfnamefont{D.}~\bibnamefont{{Merritt}}},
  \bibinfo{journal}{Modern Physics Letters A} \textbf{\bibinfo{volume}{20}},
  \bibinfo{pages}{1021} (\bibinfo{year}{2005}).

\bibitem[{\citenamefont{{Peebles}}(1972)}]{peebles-72}
\bibinfo{author}{\bibfnamefont{P.~J.~E.} \bibnamefont{{Peebles}}},
  \bibinfo{journal}{Astrophys. J.} \textbf{\bibinfo{volume}{178}},
  \bibinfo{pages}{371} (\bibinfo{year}{1972}).

\bibitem[{\citenamefont{{Young}}(1980)}]{young-80}
\bibinfo{author}{\bibfnamefont{P.}~\bibnamefont{{Young}}},
  \bibinfo{journal}{Astrophys. J.} \textbf{\bibinfo{volume}{242}},
  \bibinfo{pages}{1232} (\bibinfo{year}{1980}).

\bibitem[{\citenamefont{{Gondolo} and {Silk}}(1999)}]{GS-99}
\bibinfo{author}{\bibfnamefont{P.}~\bibnamefont{{Gondolo}}} \bibnamefont{and}
  \bibinfo{author}{\bibfnamefont{J.}~\bibnamefont{{Silk}}},
  \bibinfo{journal}{Physical Review Letters} \textbf{\bibinfo{volume}{83}},
  \bibinfo{pages}{1719} (\bibinfo{year}{1999}).

\bibitem[{\citenamefont{{Ullio} et~al.}(2001)\citenamefont{{Ullio}, {Zhao}, and
  {Kamionkowski}}}]{ullio-01}
\bibinfo{author}{\bibfnamefont{P.}~\bibnamefont{{Ullio}}},
  \bibinfo{author}{\bibfnamefont{H.}~\bibnamefont{{Zhao}}}, \bibnamefont{and}
  \bibinfo{author}{\bibfnamefont{M.}~\bibnamefont{{Kamionkowski}}},
  \bibinfo{journal}{Phys. Rev. D} \textbf{\bibinfo{volume}{64}},
  \bibinfo{pages}{043504} (\bibinfo{year}{2001}),
  \eprint{arXiv:astro-ph/0101481}.

\bibitem[{\citenamefont{{Shlosman} et~al.}(1989)\citenamefont{{Shlosman},
  {Frank}, and {Begelman}}}]{shlosman-89}
\bibinfo{author}{\bibfnamefont{I.}~\bibnamefont{{Shlosman}}},
  \bibinfo{author}{\bibfnamefont{J.}~\bibnamefont{{Frank}}}, \bibnamefont{and}
  \bibinfo{author}{\bibfnamefont{M.~C.} \bibnamefont{{Begelman}}},
  \bibinfo{journal}{Nature} \textbf{\bibinfo{volume}{338}}, \bibinfo{pages}{45}
  (\bibinfo{year}{1989}).

\bibitem[{\citenamefont{{Merritt} et~al.}(2002)\citenamefont{{Merritt},
  {Milosavljevi{\' c}}, {Verde}, and {Jimenez}}}]{MMVJ-02}
\bibinfo{author}{\bibfnamefont{D.}~\bibnamefont{{Merritt}}},
  \bibinfo{author}{\bibfnamefont{M.}~\bibnamefont{{Milosavljevi{\' c}}}},
  \bibinfo{author}{\bibfnamefont{L.}~\bibnamefont{{Verde}}}, \bibnamefont{and}
  \bibinfo{author}{\bibfnamefont{R.}~\bibnamefont{{Jimenez}}},
  \bibinfo{journal}{Phys. Rev. Lett.} \textbf{\bibinfo{volume}{88}},
  \bibinfo{pages}{191301} (\bibinfo{year}{2002}).

\bibitem[{\citenamefont{{Kauffmann} and {Haehnelt}}(2000)}]{HK-00}
\bibinfo{author}{\bibfnamefont{G.}~\bibnamefont{{Kauffmann}}} \bibnamefont{and}
  \bibinfo{author}{\bibfnamefont{M.}~\bibnamefont{{Haehnelt}}},
  \bibinfo{journal}{Mon. Not. R. Astron. Soc.} \textbf{\bibinfo{volume}{311}},
  \bibinfo{pages}{576} (\bibinfo{year}{2000}).

\bibitem[{\citenamefont{{Begelman} et~al.}(1980)\citenamefont{{Begelman},
  {Blandford}, and {Rees}}}]{BBR-80}
\bibinfo{author}{\bibfnamefont{M.~C.} \bibnamefont{{Begelman}}},
  \bibinfo{author}{\bibfnamefont{R.~D.} \bibnamefont{{Blandford}}},
  \bibnamefont{and} \bibinfo{author}{\bibfnamefont{M.~J.}
  \bibnamefont{{Rees}}}, \bibinfo{journal}{Nature}
  \textbf{\bibinfo{volume}{287}}, \bibinfo{pages}{307} (\bibinfo{year}{1980}).

\bibitem[{\citenamefont{{Saslaw} et~al.}(1974)\citenamefont{{Saslaw},
  {Valtonen}, and {Aarseth}}}]{SVA-74}
\bibinfo{author}{\bibfnamefont{W.~C.} \bibnamefont{{Saslaw}}},
  \bibinfo{author}{\bibfnamefont{M.~J.} \bibnamefont{{Valtonen}}},
  \bibnamefont{and} \bibinfo{author}{\bibfnamefont{S.~J.}
  \bibnamefont{{Aarseth}}}, \bibinfo{journal}{Astrophys. J.}
  \textbf{\bibinfo{volume}{190}}, \bibinfo{pages}{253} (\bibinfo{year}{1974}).

\bibitem[{\citenamefont{{Quinlan}}(1996)}]{quinlan-96}
\bibinfo{author}{\bibfnamefont{G.~D.} \bibnamefont{{Quinlan}}},
  \bibinfo{journal}{New Astronomy} \textbf{\bibinfo{volume}{1}},
  \bibinfo{pages}{35} (\bibinfo{year}{1996}).

\bibitem[{\citenamefont{{Graham}}(2004)}]{graham-04}
\bibinfo{author}{\bibfnamefont{A.~W.} \bibnamefont{{Graham}}},
  \bibinfo{journal}{Astrophys. J. (Lett.)} \textbf{\bibinfo{volume}{613}},
  \bibinfo{pages}{L33} (\bibinfo{year}{2004}).

\bibitem[{\citenamefont{{Volonteri} et~al.}(2003)\citenamefont{{Volonteri},
  {Madau}, and {Haardt}}}]{volonteri-03}
\bibinfo{author}{\bibfnamefont{M.}~\bibnamefont{{Volonteri}}},
  \bibinfo{author}{\bibfnamefont{P.}~\bibnamefont{{Madau}}}, \bibnamefont{and}
  \bibinfo{author}{\bibfnamefont{F.}~\bibnamefont{{Haardt}}},
  \bibinfo{journal}{Astrophys. J.} \textbf{\bibinfo{volume}{593}},
  \bibinfo{pages}{661} (\bibinfo{year}{2003}).

\bibitem[{\citenamefont{{Redmount} and {Rees}}(1989)}]{RR-89}
\bibinfo{author}{\bibfnamefont{I.~H.} \bibnamefont{{Redmount}}}
  \bibnamefont{and} \bibinfo{author}{\bibfnamefont{M.~J.}
  \bibnamefont{{Rees}}}, \bibinfo{journal}{Comments on Astrophysics}
  \textbf{\bibinfo{volume}{14}}, \bibinfo{pages}{165} (\bibinfo{year}{1989}).

\bibitem[{\citenamefont{{Campanelli} et~al.}(2007)\citenamefont{{Campanelli},
  {Lousto}, {Zlochower}, and {Merritt}}}]{lousto-07}
\bibinfo{author}{\bibfnamefont{M.}~\bibnamefont{{Campanelli}}},
  \bibinfo{author}{\bibfnamefont{C.~O.} \bibnamefont{{Lousto}}},
  \bibinfo{author}{\bibfnamefont{Y.}~\bibnamefont{{Zlochower}}},
  \bibnamefont{and}
  \bibinfo{author}{\bibfnamefont{D.}~\bibnamefont{{Merritt}}},
  \bibinfo{journal}{Physical Review Letters} \textbf{\bibinfo{volume}{98}},
  \bibinfo{pages}{231102} (\bibinfo{year}{2007}), \eprint{arXiv:gr-qc/0702133}.

\bibitem[{\citenamefont{{Merritt} et~al.}(2004)\citenamefont{{Merritt},
  {Milosavljevi{\' c}}, {Favata}, {Hughes}, and {Holz}}}]{mmfhh-04}
\bibinfo{author}{\bibfnamefont{D.}~\bibnamefont{{Merritt}}},
  \bibinfo{author}{\bibfnamefont{M.}~\bibnamefont{{Milosavljevi{\' c}}}},
  \bibinfo{author}{\bibfnamefont{M.}~\bibnamefont{{Favata}}},
  \bibinfo{author}{\bibfnamefont{S.~A.} \bibnamefont{{Hughes}}},
  \bibnamefont{and} \bibinfo{author}{\bibfnamefont{D.~E.}
  \bibnamefont{{Holz}}}, \bibinfo{journal}{Astrophys. J. (Lett.)}
  \textbf{\bibinfo{volume}{607}}, \bibinfo{pages}{L9} (\bibinfo{year}{2004}).

\bibitem[{\citenamefont{{Boylan-Kolchin}
  et~al.}(2004)\citenamefont{{Boylan-Kolchin}, {Ma}, and
  {Quataert}}}]{copycats-04}
\bibinfo{author}{\bibfnamefont{M.}~\bibnamefont{{Boylan-Kolchin}}},
  \bibinfo{author}{\bibfnamefont{C.-P.} \bibnamefont{{Ma}}}, \bibnamefont{and}
  \bibinfo{author}{\bibfnamefont{E.}~\bibnamefont{{Quataert}}},
  \bibinfo{journal}{Astrophys. J.} \textbf{\bibinfo{volume}{613}},
  \bibinfo{pages}{L37} (\bibinfo{year}{2004}).

\bibitem[{\citenamefont{{Lauer} et~al.}(2007)\citenamefont{{Lauer}, {Faber},
  {Richstone}, {Gebhardt}, {Tremaine}, {Postman}, {Dressler}, {Aller},
  {Filippenko}, {Green} et~al.}}]{lauer-07}
\bibinfo{author}{\bibfnamefont{T.~R.} \bibnamefont{{Lauer}}},
  \bibinfo{author}{\bibfnamefont{S.~M.} \bibnamefont{{Faber}}},
  \bibinfo{author}{\bibfnamefont{D.}~\bibnamefont{{Richstone}}},
  \bibinfo{author}{\bibfnamefont{K.}~\bibnamefont{{Gebhardt}}},
  \bibinfo{author}{\bibfnamefont{S.}~\bibnamefont{{Tremaine}}},
  \bibinfo{author}{\bibfnamefont{M.}~\bibnamefont{{Postman}}},
  \bibinfo{author}{\bibfnamefont{A.}~\bibnamefont{{Dressler}}},
  \bibinfo{author}{\bibfnamefont{M.~C.} \bibnamefont{{Aller}}},
  \bibinfo{author}{\bibfnamefont{A.~V.} \bibnamefont{{Filippenko}}},
  \bibinfo{author}{\bibfnamefont{R.}~\bibnamefont{{Green}}},
  \bibnamefont{et~al.}, \bibinfo{journal}{Astrophys. J.}
  \textbf{\bibinfo{volume}{662}}, \bibinfo{pages}{808} (\bibinfo{year}{2007}),
  \eprint{arXiv:astro-ph/0606739}.

\bibitem[{\citenamefont{{Merritt} and {Szell}}(2006)}]{MS-06}
\bibinfo{author}{\bibfnamefont{D.}~\bibnamefont{{Merritt}}} \bibnamefont{and}
  \bibinfo{author}{\bibfnamefont{A.}~\bibnamefont{{Szell}}},
  \bibinfo{journal}{Astrophys. J.} \textbf{\bibinfo{volume}{648}},
  \bibinfo{pages}{890} (\bibinfo{year}{2006}), \eprint{arXiv:astro-ph/0510498}.

\bibitem[{\citenamefont{{Lauer} et~al.}(1998)\citenamefont{{Lauer}, {Faber},
  {Ajhar}, {Grillmair}, and {Scowen}}}]{lauer-98}
\bibinfo{author}{\bibfnamefont{T.~R.} \bibnamefont{{Lauer}}},
  \bibinfo{author}{\bibfnamefont{S.~M.} \bibnamefont{{Faber}}},
  \bibinfo{author}{\bibfnamefont{E.~A.} \bibnamefont{{Ajhar}}},
  \bibinfo{author}{\bibfnamefont{C.~J.} \bibnamefont{{Grillmair}}},
  \bibnamefont{and} \bibinfo{author}{\bibfnamefont{P.~A.}
  \bibnamefont{{Scowen}}}, \bibinfo{journal}{Astron. J.}
  \textbf{\bibinfo{volume}{116}}, \bibinfo{pages}{2263} (\bibinfo{year}{1998}).

\bibitem[{\citenamefont{{Ilyin} et~al.}(2004)\citenamefont{{Ilyin}, {Zybin},
  and {Gurevich}}}]{ilyin-04}
\bibinfo{author}{\bibfnamefont{A.~S.} \bibnamefont{{Ilyin}}},
  \bibinfo{author}{\bibfnamefont{K.~P.} \bibnamefont{{Zybin}}},
  \bibnamefont{and} \bibinfo{author}{\bibfnamefont{A.~V.}
  \bibnamefont{{Gurevich}}}, \bibinfo{journal}{Soviet Journal of Experimental
  and Theoretical Physics} \textbf{\bibinfo{volume}{98}}, \bibinfo{pages}{1}
  (\bibinfo{year}{2004}), \eprint{arXiv:astro-ph/0306490}.

\bibitem[{\citenamefont{{Merritt}}(2004)}]{merritt-04}
\bibinfo{author}{\bibfnamefont{D.}~\bibnamefont{{Merritt}}},
  \bibinfo{journal}{Physical Review Letters} \textbf{\bibinfo{volume}{92}},
  \bibinfo{pages}{201304} (\bibinfo{year}{2004}).

\bibitem[{\citenamefont{{Merritt}
  et~al.}(2007{\natexlab{b}})\citenamefont{{Merritt}, {Harfst}, and
  {Bertone}}}]{MHB-07}
\bibinfo{author}{\bibfnamefont{D.}~\bibnamefont{{Merritt}}},
  \bibinfo{author}{\bibfnamefont{S.}~\bibnamefont{{Harfst}}}, \bibnamefont{and}
  \bibinfo{author}{\bibfnamefont{G.}~\bibnamefont{{Bertone}}},
  \bibinfo{journal}{Phys. Rev. D} \textbf{\bibinfo{volume}{75}},
  \bibinfo{pages}{043517} (\bibinfo{year}{2007}{\natexlab{b}}),
  \eprint{arXiv:astro-ph/0610425}.

\bibitem[{\citenamefont{{Baumgardt} et~al.}(2004)\citenamefont{{Baumgardt},
  {Makino}, and {Ebisuzaki}}}]{baumgardt-04b}
\bibinfo{author}{\bibfnamefont{H.}~\bibnamefont{{Baumgardt}}},
  \bibinfo{author}{\bibfnamefont{J.}~\bibnamefont{{Makino}}}, \bibnamefont{and}
  \bibinfo{author}{\bibfnamefont{T.}~\bibnamefont{{Ebisuzaki}}},
  \bibinfo{journal}{Astrophys. J.} \textbf{\bibinfo{volume}{613}},
  \bibinfo{pages}{1143} (\bibinfo{year}{2004}).

\bibitem[{\citenamefont{{Kim} et~al.}(2004)\citenamefont{{Kim}, {Figer}, and
  {Morris}}}]{Kim-04}
\bibinfo{author}{\bibfnamefont{S.~S.} \bibnamefont{{Kim}}},
  \bibinfo{author}{\bibfnamefont{D.~F.} \bibnamefont{{Figer}}},
  \bibnamefont{and} \bibinfo{author}{\bibfnamefont{M.}~\bibnamefont{{Morris}}},
  \bibinfo{journal}{Astrophys. J. (Lett.)} \textbf{\bibinfo{volume}{607}},
  \bibinfo{pages}{L123} (\bibinfo{year}{2004}).

\bibitem[{\citenamefont{{Stecker}}(1988)}]{Stecker-88}
\bibinfo{author}{\bibfnamefont{F.~W.} \bibnamefont{{Stecker}}},
  \bibinfo{journal}{Physics Letters B} \textbf{\bibinfo{volume}{201}},
  \bibinfo{pages}{529} (\bibinfo{year}{1988}).

\bibitem[{\citenamefont{{Bergstr{\"o}m}
  et~al.}(1998)\citenamefont{{Bergstr{\"o}m}, {Ullio}, and {Buckley}}}]{BUB-98}
\bibinfo{author}{\bibfnamefont{L.}~\bibnamefont{{Bergstr{\"o}m}}},
  \bibinfo{author}{\bibfnamefont{P.}~\bibnamefont{{Ullio}}}, \bibnamefont{and}
  \bibinfo{author}{\bibfnamefont{J.~H.} \bibnamefont{{Buckley}}},
  \bibinfo{journal}{Astroparticle Physics} \textbf{\bibinfo{volume}{9}},
  \bibinfo{pages}{137} (\bibinfo{year}{1998}), \eprint{arXiv:astro-ph/9712318}.

\bibitem[{\citenamefont{{Bertone} et~al.}(2001)\citenamefont{{Bertone}, {Sigl},
  and {Silk}}}]{BSS-01}
\bibinfo{author}{\bibfnamefont{G.}~\bibnamefont{{Bertone}}},
  \bibinfo{author}{\bibfnamefont{G.}~\bibnamefont{{Sigl}}}, \bibnamefont{and}
  \bibinfo{author}{\bibfnamefont{J.}~\bibnamefont{{Silk}}},
  \bibinfo{journal}{Mon. Not. R. Astron. Soc.} \textbf{\bibinfo{volume}{326}},
  \bibinfo{pages}{799} (\bibinfo{year}{2001}), \eprint{arXiv:astro-ph/0101134}.

\bibitem[{\citenamefont{{Trippe} et~al.}(2008)\citenamefont{{Trippe},
  {Gillessen}, {Gerhard}, {Bartko}, {Fritz}, {Maness}, {Eisenhauer}, {Martins},
  {Ott}, {Dodds-Eden} et~al.}}]{trippe-08}
\bibinfo{author}{\bibfnamefont{S.}~\bibnamefont{{Trippe}}},
  \bibinfo{author}{\bibfnamefont{S.}~\bibnamefont{{Gillessen}}},
  \bibinfo{author}{\bibfnamefont{O.~E.} \bibnamefont{{Gerhard}}},
  \bibinfo{author}{\bibfnamefont{H.}~\bibnamefont{{Bartko}}},
  \bibinfo{author}{\bibfnamefont{T.~K.} \bibnamefont{{Fritz}}},
  \bibinfo{author}{\bibfnamefont{H.~L.} \bibnamefont{{Maness}}},
  \bibinfo{author}{\bibfnamefont{F.}~\bibnamefont{{Eisenhauer}}},
  \bibinfo{author}{\bibfnamefont{F.}~\bibnamefont{{Martins}}},
  \bibinfo{author}{\bibfnamefont{T.}~\bibnamefont{{Ott}}},
  \bibinfo{author}{\bibfnamefont{K.}~\bibnamefont{{Dodds-Eden}}},
  \bibnamefont{et~al.}, \bibinfo{journal}{ArXiv e-prints}
  (\bibinfo{year}{2008}), \eprint{0810.1040}.

\bibitem[{\citenamefont{{Sch{\"o}del}
  et~al.}(2008{\natexlab{b}})\citenamefont{{Sch{\"o}del}, {Merritt}, and
  {Eckart}}}]{pmpaper-08}
\bibinfo{author}{\bibfnamefont{R.}~\bibnamefont{{Sch{\"o}del}}},
  \bibinfo{author}{\bibfnamefont{D.}~\bibnamefont{{Merritt}}},
  \bibnamefont{and} \bibinfo{author}{\bibfnamefont{A.}~\bibnamefont{{Eckart}}},
  \bibinfo{journal}{Journal of Physics Conference Series}
  \textbf{\bibinfo{volume}{131}}, \bibinfo{pages}{012044}
  (\bibinfo{year}{2008}{\natexlab{b}}), \eprint{0810.0204}.

\bibitem[{\citenamefont{{Hall} and {Gondolo}}(2006)}]{hall-06}
\bibinfo{author}{\bibfnamefont{J.}~\bibnamefont{{Hall}}} \bibnamefont{and}
  \bibinfo{author}{\bibfnamefont{P.}~\bibnamefont{{Gondolo}}},
  \bibinfo{journal}{Phys. Rev. D} \textbf{\bibinfo{volume}{74}},
  \bibinfo{pages}{063511} (\bibinfo{year}{2006}),
  \eprint{arXiv:astro-ph/0602400}.

\bibitem[{\citenamefont{{Zakharov} et~al.}(2007)\citenamefont{{Zakharov},
  {Nucita}, {de Paolis}, and {Ingrosso}}}]{zakharov-08}
\bibinfo{author}{\bibfnamefont{A.~F.} \bibnamefont{{Zakharov}}},
  \bibinfo{author}{\bibfnamefont{A.~A.} \bibnamefont{{Nucita}}},
  \bibinfo{author}{\bibfnamefont{F.}~\bibnamefont{{de Paolis}}},
  \bibnamefont{and}
  \bibinfo{author}{\bibfnamefont{G.}~\bibnamefont{{Ingrosso}}},
  \bibinfo{journal}{Physical Review D} \textbf{\bibinfo{volume}{76}},
  \bibinfo{pages}{062001} (\bibinfo{year}{2007}), \eprint{0707.4423}.

\bibitem[{\citenamefont{{Hofmann} and {The Hess
  Collaboration}}(2000)}]{HESS-00}
\bibinfo{author}{\bibfnamefont{W.}~\bibnamefont{{Hofmann}}} \bibnamefont{and}
  \bibinfo{author}{\bibnamefont{{The Hess Collaboration}}}, in
  \emph{\bibinfo{booktitle}{American Institute of Physics Conference Series}},
  edited by \bibinfo{editor}{\bibfnamefont{B.~L.} \bibnamefont{{Dingus}}},
  \bibinfo{editor}{\bibfnamefont{M.~H.} \bibnamefont{{Salamon}}},
  \bibnamefont{and} \bibinfo{editor}{\bibfnamefont{D.~B.}
  \bibnamefont{{Kieda}}} (\bibinfo{year}{2000}), vol. \bibinfo{volume}{515} of
  \emph{\bibinfo{series}{American Institute of Physics Conference Series}}, pp.
  \bibinfo{pages}{500--+}.

\bibitem[{\citenamefont{{Moiseev}}(2008)}]{GLAST}
\bibinfo{author}{\bibfnamefont{A.~A.} \bibnamefont{{Moiseev}}},
  \bibinfo{journal}{Nuclear Instruments and Methods in Physics Research A}
  \textbf{\bibinfo{volume}{588}}, \bibinfo{pages}{41} (\bibinfo{year}{2008}).

\bibitem[{\citenamefont{{Kosack}}(2004)}]{Whipple-04}
\bibinfo{author}{\bibfnamefont{K.~e.~a.} \bibnamefont{{Kosack}}},
  \bibinfo{journal}{Astrophys. J. Letts.} \textbf{\bibinfo{volume}{608}},
  \bibinfo{pages}{L97} (\bibinfo{year}{2004}), \eprint{arXiv:astro-ph/0403422}.

\bibitem[{\citenamefont{{Aharonian}}(2004)}]{HESS-04}
\bibinfo{author}{\bibfnamefont{F.~e.~a.} \bibnamefont{{Aharonian}}},
  \bibinfo{journal}{Astron. Astrophys.} \textbf{\bibinfo{volume}{425}},
  \bibinfo{pages}{L13} (\bibinfo{year}{2004}), \eprint{arXiv:astro-ph/0406658}.

\bibitem[{\citenamefont{{Hooper} et~al.}(2004)\citenamefont{{Hooper}, {de la
  Calle Perez}, {Silk}, {Ferrer}, and {Sarkar}}}]{hooper-04}
\bibinfo{author}{\bibfnamefont{D.}~\bibnamefont{{Hooper}}},
  \bibinfo{author}{\bibfnamefont{I.}~\bibnamefont{{de la Calle Perez}}},
  \bibinfo{author}{\bibfnamefont{J.}~\bibnamefont{{Silk}}},
  \bibinfo{author}{\bibfnamefont{F.}~\bibnamefont{{Ferrer}}}, \bibnamefont{and}
  \bibinfo{author}{\bibfnamefont{S.}~\bibnamefont{{Sarkar}}},
  \bibinfo{journal}{Journal of Cosmology and Astro-Particle Physics}
  \textbf{\bibinfo{volume}{9}}, \bibinfo{pages}{2} (\bibinfo{year}{2004}),
  \eprint{arXiv:astro-ph/0404205}.

\bibitem[{\citenamefont{{Horns}}(2005)}]{horns-05}
\bibinfo{author}{\bibfnamefont{D.}~\bibnamefont{{Horns}}},
  \bibinfo{journal}{Physics Letters B} \textbf{\bibinfo{volume}{607}},
  \bibinfo{pages}{225} (\bibinfo{year}{2005}), \eprint{arXiv:astro-ph/0408192}.

\bibitem[{\citenamefont{{Profumo}}(2005)}]{profumo-05}
\bibinfo{author}{\bibfnamefont{S.}~\bibnamefont{{Profumo}}},
  \bibinfo{journal}{Phys. Rev. D} \textbf{\bibinfo{volume}{72}},
  \bibinfo{pages}{103521} (\bibinfo{year}{2005}),
  \eprint{arXiv:astro-ph/0508628}.

\bibitem[{\citenamefont{{Vasiliev} and {Zelnikov}}(2008)}]{vasiliev-08}
\bibinfo{author}{\bibfnamefont{E.}~\bibnamefont{{Vasiliev}}} \bibnamefont{and}
  \bibinfo{author}{\bibfnamefont{M.}~\bibnamefont{{Zelnikov}}},
  \bibinfo{journal}{Physical Review D} \textbf{\bibinfo{volume}{78}},
  \bibinfo{pages}{083506} (\bibinfo{year}{2008}), \eprint{0803.0002}.

\bibitem[{\citenamefont{{Berezinsky} et~al.}(1992)\citenamefont{{Berezinsky},
  {Gurevich}, and {Zybin}}}]{berezinsky-92}
\bibinfo{author}{\bibfnamefont{V.~S.} \bibnamefont{{Berezinsky}}},
  \bibinfo{author}{\bibfnamefont{A.~V.} \bibnamefont{{Gurevich}}},
  \bibnamefont{and} \bibinfo{author}{\bibfnamefont{K.~P.}
  \bibnamefont{{Zybin}}}, \bibinfo{journal}{Physics Letters B}
  \textbf{\bibinfo{volume}{294}}, \bibinfo{pages}{221} (\bibinfo{year}{1992}).

\bibitem[{\citenamefont{{Vasiliev}}(2007)}]{vasiliev-07}
\bibinfo{author}{\bibfnamefont{E.}~\bibnamefont{{Vasiliev}}},
  \bibinfo{journal}{Physical Review D} \textbf{\bibinfo{volume}{76}},
  \bibinfo{pages}{103532} (\bibinfo{year}{2007}), \eprint{0707.3334}.

\bibitem[{\citenamefont{{Eidelman}}(2004)}]{eidelman-04}
\bibinfo{author}{\bibfnamefont{S.~e.~a.} \bibnamefont{{Eidelman}}},
  \bibinfo{journal}{{Physics Letters B}} \textbf{\bibinfo{volume}{592}},
  \bibinfo{pages}{1+} (\bibinfo{year}{2004}),
  \urlprefix\url{http://pdg.lbl.gov}.

\bibitem[{\citenamefont{{Berezinsky} et~al.}(1994)\citenamefont{{Berezinsky},
  {Bottino}, and {Mignola}}}]{berezinsky-94}
\bibinfo{author}{\bibfnamefont{V.}~\bibnamefont{{Berezinsky}}},
  \bibinfo{author}{\bibfnamefont{A.}~\bibnamefont{{Bottino}}},
  \bibnamefont{and}
  \bibinfo{author}{\bibfnamefont{G.}~\bibnamefont{{Mignola}}},
  \bibinfo{journal}{Physics Letters B} \textbf{\bibinfo{volume}{325}},
  \bibinfo{pages}{136} (\bibinfo{year}{1994}), \eprint{arXiv:hep-ph/9402215}.

\bibitem[{\citenamefont{{Bertone}}(2007)}]{bertone-07}
\bibinfo{author}{\bibfnamefont{G.}~\bibnamefont{{Bertone}}},
  \bibinfo{journal}{Astrophys. Sp. Sci.} \textbf{\bibinfo{volume}{309}},
  \bibinfo{pages}{505} (\bibinfo{year}{2007}), \eprint{arXiv:astro-ph/0608706}.

\bibitem[{\citenamefont{{Forbes} et~al.}(2008)\citenamefont{{Forbes}, {Lasky},
  {Graham}, and {Spitler}}}]{forbes-08}
\bibinfo{author}{\bibfnamefont{D.~A.} \bibnamefont{{Forbes}}},
  \bibinfo{author}{\bibfnamefont{P.}~\bibnamefont{{Lasky}}},
  \bibinfo{author}{\bibfnamefont{A.~W.} \bibnamefont{{Graham}}},
  \bibnamefont{and}
  \bibinfo{author}{\bibfnamefont{L.}~\bibnamefont{{Spitler}}},
  \bibinfo{journal}{Mon. Not. R. Astron. Soc.} \textbf{\bibinfo{volume}{389}},
  \bibinfo{pages}{1924} (\bibinfo{year}{2008}), \eprint{0806.1090}.

\bibitem[{\citenamefont{{Mateo}}(1998)}]{mateo-98}
\bibinfo{author}{\bibfnamefont{M.~L.} \bibnamefont{{Mateo}}},
  \bibinfo{journal}{Ann. Rev. Astron. Astrophys.}
  \textbf{\bibinfo{volume}{36}}, \bibinfo{pages}{435} (\bibinfo{year}{1998}),
  \eprint{arXiv:astro-ph/9810070}.

\bibitem[{\citenamefont{{Merritt}}(1993)}]{merritt-93}
\bibinfo{author}{\bibfnamefont{D.}~\bibnamefont{{Merritt}}},
  \bibinfo{journal}{Astrophys. J.} \textbf{\bibinfo{volume}{413}},
  \bibinfo{pages}{79} (\bibinfo{year}{1993}).

\bibitem[{\citenamefont{{Kleyna} et~al.}(2002)\citenamefont{{Kleyna},
  {Wilkinson}, {Evans}, {Gilmore}, and {Frayn}}}]{kleyna-02}
\bibinfo{author}{\bibfnamefont{J.}~\bibnamefont{{Kleyna}}},
  \bibinfo{author}{\bibfnamefont{M.~I.} \bibnamefont{{Wilkinson}}},
  \bibinfo{author}{\bibfnamefont{N.~W.} \bibnamefont{{Evans}}},
  \bibinfo{author}{\bibfnamefont{G.}~\bibnamefont{{Gilmore}}},
  \bibnamefont{and} \bibinfo{author}{\bibfnamefont{C.}~\bibnamefont{{Frayn}}},
  \bibinfo{journal}{Monthly Notices of the Royal Astronomical Society}
  \textbf{\bibinfo{volume}{330}}, \bibinfo{pages}{792} (\bibinfo{year}{2002}),
  \eprint{arXiv:astro-ph/0109450}.

\bibitem[{\citenamefont{{Mashchenko} et~al.}(2006)\citenamefont{{Mashchenko},
  {Sills}, and {Couchman}}}]{masch-06}
\bibinfo{author}{\bibfnamefont{S.}~\bibnamefont{{Mashchenko}}},
  \bibinfo{author}{\bibfnamefont{A.}~\bibnamefont{{Sills}}}, \bibnamefont{and}
  \bibinfo{author}{\bibfnamefont{H.~M.} \bibnamefont{{Couchman}}},
  \bibinfo{journal}{The Astrophysical Journal} \textbf{\bibinfo{volume}{640}},
  \bibinfo{pages}{252} (\bibinfo{year}{2006}), \eprint{arXiv:astro-ph/0511567}.

\bibitem[{\citenamefont{{Walker} et~al.}(2007)\citenamefont{{Walker}, {Mateo},
  {Olszewski}, {Gnedin}, {Wang}, {Sen}, and {Woodroofe}}}]{walker-07}
\bibinfo{author}{\bibfnamefont{M.~G.} \bibnamefont{{Walker}}},
  \bibinfo{author}{\bibfnamefont{M.}~\bibnamefont{{Mateo}}},
  \bibinfo{author}{\bibfnamefont{E.~W.} \bibnamefont{{Olszewski}}},
  \bibinfo{author}{\bibfnamefont{O.~Y.} \bibnamefont{{Gnedin}}},
  \bibinfo{author}{\bibfnamefont{X.}~\bibnamefont{{Wang}}},
  \bibinfo{author}{\bibfnamefont{B.}~\bibnamefont{{Sen}}}, \bibnamefont{and}
  \bibinfo{author}{\bibfnamefont{M.}~\bibnamefont{{Woodroofe}}},
  \bibinfo{journal}{The Astrophysical Journal Letters}
  \textbf{\bibinfo{volume}{667}}, \bibinfo{pages}{L53} (\bibinfo{year}{2007}),
  \eprint{0708.0010}.

\bibitem[{\citenamefont{{Kleyna} et~al.}(2003)\citenamefont{{Kleyna},
  {Wilkinson}, {Gilmore}, and {Evans}}}]{kleyna-03}
\bibinfo{author}{\bibfnamefont{J.~T.} \bibnamefont{{Kleyna}}},
  \bibinfo{author}{\bibfnamefont{M.~I.} \bibnamefont{{Wilkinson}}},
  \bibinfo{author}{\bibfnamefont{G.}~\bibnamefont{{Gilmore}}},
  \bibnamefont{and} \bibinfo{author}{\bibfnamefont{N.~W.}
  \bibnamefont{{Evans}}}, \bibinfo{journal}{Astrophys. J. Letts.}
  \textbf{\bibinfo{volume}{588}}, \bibinfo{pages}{L21} (\bibinfo{year}{2003}),
  \eprint{arXiv:astro-ph/0304093}.

\bibitem[{\citenamefont{{Goerdt} et~al.}(2006)\citenamefont{{Goerdt}, {Moore},
  {Read}, {Stadel}, and {Zemp}}}]{goerdt-06}
\bibinfo{author}{\bibfnamefont{T.}~\bibnamefont{{Goerdt}}},
  \bibinfo{author}{\bibfnamefont{B.}~\bibnamefont{{Moore}}},
  \bibinfo{author}{\bibfnamefont{J.~I.} \bibnamefont{{Read}}},
  \bibinfo{author}{\bibfnamefont{J.}~\bibnamefont{{Stadel}}}, \bibnamefont{and}
  \bibinfo{author}{\bibfnamefont{M.}~\bibnamefont{{Zemp}}},
  \bibinfo{journal}{Mon. Not. R. Astron. Soc.} \textbf{\bibinfo{volume}{368}},
  \bibinfo{pages}{1073} (\bibinfo{year}{2006}),
  \eprint{arXiv:astro-ph/0601404}.

\bibitem[{\citenamefont{{Gilmore} et~al.}(2007)\citenamefont{{Gilmore},
  {Wilkinson}, {Wyse}, {Kleyna}, {Koch}, {Evans}, and {Grebel}}}]{gilmore-07}
\bibinfo{author}{\bibfnamefont{G.}~\bibnamefont{{Gilmore}}},
  \bibinfo{author}{\bibfnamefont{M.~I.} \bibnamefont{{Wilkinson}}},
  \bibinfo{author}{\bibfnamefont{R.~F.~G.} \bibnamefont{{Wyse}}},
  \bibinfo{author}{\bibfnamefont{J.~T.} \bibnamefont{{Kleyna}}},
  \bibinfo{author}{\bibfnamefont{A.}~\bibnamefont{{Koch}}},
  \bibinfo{author}{\bibfnamefont{N.~W.} \bibnamefont{{Evans}}},
  \bibnamefont{and} \bibinfo{author}{\bibfnamefont{E.~K.}
  \bibnamefont{{Grebel}}}, \bibinfo{journal}{Astrophys. J.}
  \textbf{\bibinfo{volume}{663}}, \bibinfo{pages}{948} (\bibinfo{year}{2007}),
  \eprint{arXiv:astro-ph/0703308}.

\bibitem[{\citenamefont{{Belokurov} et~al.}(2007)\citenamefont{{Belokurov},
  {Zucker}, {Evans}, {Kleyna}, {Koposov}, {Hodgkin}, {Irwin}, {Gilmore},
  {Wilkinson}, {Fellhauer} et~al.}}]{belo-07}
\bibinfo{author}{\bibfnamefont{V.}~\bibnamefont{{Belokurov}}},
  \bibinfo{author}{\bibfnamefont{D.~B.} \bibnamefont{{Zucker}}},
  \bibinfo{author}{\bibfnamefont{N.~W.} \bibnamefont{{Evans}}},
  \bibinfo{author}{\bibfnamefont{J.~T.} \bibnamefont{{Kleyna}}},
  \bibinfo{author}{\bibfnamefont{S.}~\bibnamefont{{Koposov}}},
  \bibinfo{author}{\bibfnamefont{S.~T.} \bibnamefont{{Hodgkin}}},
  \bibinfo{author}{\bibfnamefont{M.~J.} \bibnamefont{{Irwin}}},
  \bibinfo{author}{\bibfnamefont{G.}~\bibnamefont{{Gilmore}}},
  \bibinfo{author}{\bibfnamefont{M.~I.} \bibnamefont{{Wilkinson}}},
  \bibinfo{author}{\bibfnamefont{M.}~\bibnamefont{{Fellhauer}}},
  \bibnamefont{et~al.}, \bibinfo{journal}{The Astrophysical Journal}
  \textbf{\bibinfo{volume}{654}}, \bibinfo{pages}{897} (\bibinfo{year}{2007}),
  \eprint{arXiv:astro-ph/0608448}.

\bibitem[{\citenamefont{{Baltz} et~al.}(2000)\citenamefont{{Baltz}, {Briot},
  {Salati}, {Taillet}, and {Silk}}}]{baltz-00}
\bibinfo{author}{\bibfnamefont{E.~A.} \bibnamefont{{Baltz}}},
  \bibinfo{author}{\bibfnamefont{C.}~\bibnamefont{{Briot}}},
  \bibinfo{author}{\bibfnamefont{P.}~\bibnamefont{{Salati}}},
  \bibinfo{author}{\bibfnamefont{R.}~\bibnamefont{{Taillet}}},
  \bibnamefont{and} \bibinfo{author}{\bibfnamefont{J.}~\bibnamefont{{Silk}}},
  \bibinfo{journal}{Phys. Rev. D} \textbf{\bibinfo{volume}{61}},
  \bibinfo{pages}{023514} (\bibinfo{year}{2000}),
  \eprint{arXiv:astro-ph/9909112}.

\bibitem[{\citenamefont{{Strigari} et~al.}(2007)\citenamefont{{Strigari},
  {Koushiappas}, {Bullock}, and {Kaplinghat}}}]{strigari-07}
\bibinfo{author}{\bibfnamefont{L.~E.} \bibnamefont{{Strigari}}},
  \bibinfo{author}{\bibfnamefont{S.~M.} \bibnamefont{{Koushiappas}}},
  \bibinfo{author}{\bibfnamefont{J.~S.} \bibnamefont{{Bullock}}},
  \bibnamefont{and}
  \bibinfo{author}{\bibfnamefont{M.}~\bibnamefont{{Kaplinghat}}},
  \bibinfo{journal}{Phys. Rev. D} \textbf{\bibinfo{volume}{75}},
  \bibinfo{pages}{083526} (\bibinfo{year}{2007}),
  \eprint{arXiv:astro-ph/0611925}.

\bibitem[{\citenamefont{{Evans} et~al.}(2004)\citenamefont{{Evans}, {Ferrer},
  and {Sarkar}}}]{evans-04}
\bibinfo{author}{\bibfnamefont{N.~W.} \bibnamefont{{Evans}}},
  \bibinfo{author}{\bibfnamefont{F.}~\bibnamefont{{Ferrer}}}, \bibnamefont{and}
  \bibinfo{author}{\bibfnamefont{S.}~\bibnamefont{{Sarkar}}},
  \bibinfo{journal}{Physical Review D} \textbf{\bibinfo{volume}{69}},
  \bibinfo{pages}{123501} (\bibinfo{year}{2004}),
  \eprint{arXiv:astro-ph/0311145}.

\bibitem[{\citenamefont{{Strigari} et~al.}(2008)\citenamefont{{Strigari},
  {Koushiappas}, {Bullock}, {Kaplinghat}, {Simon}, {Geha}, and
  {Willman}}}]{strigari-08}
\bibinfo{author}{\bibfnamefont{L.~E.} \bibnamefont{{Strigari}}},
  \bibinfo{author}{\bibfnamefont{S.~M.} \bibnamefont{{Koushiappas}}},
  \bibinfo{author}{\bibfnamefont{J.~S.} \bibnamefont{{Bullock}}},
  \bibinfo{author}{\bibfnamefont{M.}~\bibnamefont{{Kaplinghat}}},
  \bibinfo{author}{\bibfnamefont{J.~D.} \bibnamefont{{Simon}}},
  \bibinfo{author}{\bibfnamefont{M.}~\bibnamefont{{Geha}}}, \bibnamefont{and}
  \bibinfo{author}{\bibfnamefont{B.}~\bibnamefont{{Willman}}},
  \bibinfo{journal}{Astrophys. J.} \textbf{\bibinfo{volume}{678}},
  \bibinfo{pages}{614} (\bibinfo{year}{2008}), \eprint{0709.1510}.

\bibitem[{\citenamefont{{Willman} et~al.}(2005)\citenamefont{{Willman},
  {Blanton}, {West}, {Dalcanton}, {Hogg}, {Schneider}, {Wherry}, {Yanny}, and
  {Brinkmann}}}]{willman-05}
\bibinfo{author}{\bibfnamefont{B.}~\bibnamefont{{Willman}}},
  \bibinfo{author}{\bibfnamefont{M.~R.} \bibnamefont{{Blanton}}},
  \bibinfo{author}{\bibfnamefont{A.~A.} \bibnamefont{{West}}},
  \bibinfo{author}{\bibfnamefont{J.~J.} \bibnamefont{{Dalcanton}}},
  \bibinfo{author}{\bibfnamefont{D.~W.} \bibnamefont{{Hogg}}},
  \bibinfo{author}{\bibfnamefont{D.~P.} \bibnamefont{{Schneider}}},
  \bibinfo{author}{\bibfnamefont{N.}~\bibnamefont{{Wherry}}},
  \bibinfo{author}{\bibfnamefont{B.}~\bibnamefont{{Yanny}}}, \bibnamefont{and}
  \bibinfo{author}{\bibfnamefont{J.}~\bibnamefont{{Brinkmann}}},
  \bibinfo{journal}{Astron. J.} \textbf{\bibinfo{volume}{129}},
  \bibinfo{pages}{2692} (\bibinfo{year}{2005}),
  \eprint{arXiv:astro-ph/0410416}.

\end{thebibliography}
\end{document}